\documentclass[12pt]{article}

\usepackage[english, french]{babel}
\usepackage[utf8x]{inputenc}
\usepackage{hyperref}

\usepackage[T1]{fontenc}
\usepackage[top=2.5cm, bottom=2.5cm, left=2.5cm, right=2.5cm]{geometry}
\usepackage{tabto}
\usepackage{enumitem}

\usepackage{titlesec}
\usepackage{lineno}

\titleformat{\section}[hang]
  {\normalfont\Huge\bfseries}
  {\thesection}
  {1em}
  {}

\titlespacing{\section}{0pt}{10pt plus 2pt minus 2pt}{20pt plus 1.5pt minus 0.25pt}
\usepackage{array}
\usepackage{tabularx}
\usepackage{multirow}
\usepackage[graphicx]{realboxes}
\usepackage{pgfplots}
\usepackage{pgfplotstable}

\usepackage{array}
\usepackage{tabularx}
\usepackage{ragged2e}
\usepackage{float}
\usepackage{cite}

\usepackage{float}
\usepackage{amsmath}
\usepackage{newtxtext,newtxmath}
\usepackage[justification=centering]{caption}
\usepackage[colorinlistoftodos]{todonotes}
\usepackage{url}
\counterwithout{figure}{section}

\usepackage{subcaption}
\usepackage{setspace}
\title{Photonic Interactions with Semiconducting Barrier Discharges}

\author{Ayah Soundous Taihi}
\author{David Z. Pai}

\begin{document}

%
\newcommand{\HRule}{\rule{\linewidth}{0.2mm}}


\selectlanguage{english}

{\centering

   {\begin{spacing}{1.3}
{\Huge \bfseries Photonic Interactions with \\ Semiconducting Barrier Discharges\\[0.6cm]}
    \end{spacing}}

    {\large Ayah Soundous Taihi$^{1}$, David Z. Pai$^{1*}$\\[0.1cm]}

    {\normalsize
        $^{1}$Laboratoire de Physique des Plasmas, Centre National de la Recherche Scientifique, École polytechnique, Institut Polytechnique de Paris, Sorbonne Université, Route de Saclay, 91128, Palaiseau, France.\\
       \textbf{Corresponding author:} $^{*}$\texttt{david.pai@lpp.polytechnique.fr}
    }

    \vspace{0.6em}

}

{\centering\large\bfseries Abstract\par}

\vspace{0.4em}

\noindent
Semiconducting Barrier Discharges (SeBDs) generate uniform
ionization waves in air at atmospheric pressure. In this work, we investigate how externally applied irradiation synchronized with the discharge can mimic photoconductive-type coupling between the plasma and the semiconductor surface.
By illuminating the Si-SiO$_2$ interface with nanosecond pulsed irradiation at wavelengths from 532 nm to 1064 nm, and using fast imaging, optical emission spectroscopy, and current-voltage measurements, we demonstrate that the photoexcitation of charge carriers in silicon enhances the plasma emission and increases the reduced electric field, with no detectable change in the electrical energy. The magnitude and thresholds of these responses depend on wavelength. By comparing the SeBD to a MOS photodetector, this behaviour can be explained by the absorption length. This length determines whether carriers are photogenerated inside the depletion region at the SiO$_2$-Si interface, where they are efficiently separated and undergo impact-ionization amplification, or deeper in the silicon bulk where carrier separation is weaker and free-carrier absorption diminishes the quantum efficiency.  These results focus on the microscopic processes governing the plasma-semiconductor coupling and demonstrate how the optoelectronic properties of silicon can influence surface ionization waves. 

\vspace{0.5em}



\clearpage

\section{Introduction}

Plasma-surface interaction is a core area of fundamental and applied research aimed at improving the properties of plasmas using materials, and vice versa. This is especially true for a number of atmospheric-pressure plasmas (APP) generated as surface discharges. These include dielectric barrier discharges (DBDs) \cite{leonov2014dynamics,starikovskii2009sdbd} and plasma jets \cite{viegas2022physics}, whose potential applications range from aerodynamic flow control \cite{moreau2007airflow,leonov2016dynamics} and propulsion \cite{xu2019dielectric} to medicine \cite{lin2023characterization}, surface treatment \cite{adesina2024review}, and plasma-assisted combustion \cite{zhao2025investigation}.

The properties of surface discharges depend on the properties of the materials on which they propagate. For many applications, operation in air at atmospheric pressure is necessary or highly advantageous, but in these conditions surface discharges are often neither homogeneous nor high in energy density \cite{kogelschatz2003filamentary}.
This inhomogeneity and low energy density can limit the effectiveness of processes like surface modification, where a consistent treatment is crucial \cite{fang2009polytetrafluoroethylene}. Achieving uniformity is challenging at atmospheric pressure because the discharge tends to destabilize into streamers, a class of ionization waves with a thin channel-like geometry.

Streamer formation in surface DBDs has been linked to the accumulation of surface charge, which modifies the local electric field and discharge behaviour. To delay or suppress streamer formation, several techniques have been explored. 
Some approaches involve integrating weakly conductive layers into the dielectric \cite{leonov2014dynamics,guo2009separation} to gradually drain charge between pulses or adding a third electrode \cite{sato2021surface,opaits2008surface} to extract charge from distant regions. Other strategies include adjusting the dielectric properties, specifically by lowering the permittivity or increasing the dielectric thickness \cite{zhu2017nanosecond, sokolova2019barrier, ndong2013geometrical}, which helps to reduce the surface charge density but also decreases the discharge energy. However, these previous studies have primarily investigated the type of applied voltage and electrode geometry, with comparatively less attention given to alternative materials which have largely remained dielectric types, including porous media \cite{zhang2018propagation}, ferroelectrics \cite{johnson2014ferroelectric}, pyroelectrics \cite{johnson2014thermally}, as well as liquids \cite{petrishchev2014studies}.

A promising approach to overcoming these limitations involves expanding the range of material classes used in APP reactors. Semiconductors have previously been explored for draining surface charge \cite{guo2009separation}, including using a diode array integrated into a barrier of a DBD \cite{starikovskiy2013dielectric}. Also, silicon substrates have been used in the development of plasma transistors \cite{eden2009plasma} as well as in the fabrication of microplasma arrays \cite{dussart2010integrated} and \cite{eden2005recent, eden2003microplasma}. These investigations relied on the photogeneration of electron-hole pairs in silicon induced by plasma emission, predominantly employing microcavity geometries in rare gases rather than an unconfined geometry in air typical of surface DBD studies.

Recently, Darny \textit{et al.} sought to exploit photoconductive effects with semiconductors to enhance interactions at APP interfaces \cite{darny2020uniform}. 
By using a thermal oxide silicon wafer as part of the barrier, surface semiconducting barrier discharges (SeBD) generated homogeneous ionization waves with high energy density.
Uniform plasma propagation continues along a surface throughout the entire duration of the discharge, never branching into streamers. This contrasts with DBDs, where the quasi-uniformity of surface plasmas is achievable only under certain conditions, with streamers propagating in closely packed fashion. Moreover, surface SeBDs maintain true uniformity in both positive and negative voltage polarities, whereas at atmospheric pressure surface DBDs attain quasi-uniformity only in the negative polarity.

In this previous study by Darny \textit{et al.} \cite{darny2020uniform}, illuminating the thermal oxide wafer surface with a continuous wave laser at 532 nm increased plasma emission intensity and guided the plasma toward the laser spot. This suggests a strong photoconductive coupling between the air plasma and the silicon. Subsequently, Orrière \textit{et al.} \cite{Orriere2026APL} used pulsed laser irradiation at 532 nm to demonstrate that this effect is not related to the desorption of surface charges from the SiO$_2$ layer. Other previous work on microplasma-based photodetectors and bipolar junction transistors \cite{park2002photodetection, ostrom2005microcavity, wagner2010coupling, tchertchian2011control, li2013modulating, park200540000pixel} also discussed the role of plasma photons or external irradiation in the physics of these devices. In these studies both the plasma and the photon source were continuous in time rather than pulsed. Furthermore, the photonic interaction mechanism of these microplasma devices must be distinct from that of the SeBD, as demonstrated by Orrière \textit{et al.} \cite{Orriere2026APL}.

In the present study, our primary objective is to build upon this previous work {\cite{darny2020uniform} by examining how external illumination can mimic the photonic interaction between the plasma and the semiconductor. 
To do so, we will study a SeBD is generated using nanosecond voltage pulses in atmospheric-pressure air on a Si-SiO$_2$ substrate (Section \ref{sec:plasmagen}) using diagnostics including current-voltage measurements (Section \ref{sec:plasmagen}), fast imaging, and optical emission spectroscopy (Section \ref{iccdoes}).
By directing pulsed laser illumination at various wavelengths (Section \ref{illumination}) onto the semiconductor surface, we will demonstrate how this external light influences the characterized discharge (Section \ref{discharge}). In particular, the plasma emission intensity (Section \ref{intensity}), electric field (Section \ref{efield}), and discharge energy (Section \ref{energy}) will be analyzed.
These measurements will be discussed in terms of the fate of photons penetrating the air-SiO$_2$-Si interface and silicon bulk (Sections \ref{airinterface}, \ref{bulkpenetration}, and \ref{sec:energydisc}).

\section{Experimental Setup}

The setup comprised three main components: plasma generation, fast imaging coupled with Optical Emission Spectroscopy (OES), and external light illumination. We performed three synchronized measurements to gain a comprehensive understanding of the plasma behaviour during the external light irradiation: fast imaging was used to measure plasma emission intensity, OES was employed to detect changes in the plasma electric field, and current-voltage measurements provided information on the discharge energy by measuring the total current originating from both the gas and solid phases.

\subsection{Plasma generation} \label{sec:plasmagen}

First, we begin with an overview of the discharge circuit for plasma generation. The surface SeBD was generated using the barrier discharge geometry shown in Figure \ref{2elecdiag}. A 100-µm diameter tungsten pin electrode (Figure \ref{2electrode}) was placed in mechanical contact with a thermal oxide silicon wafer (WaferNet) cut with a surface area of $1.2 \times 1.2$ cm$^2$. The wafer consisted of <100> oriented, p-doped silicon with a resistivity of $1-10$ $\Omega\cdot$cm and a thickness of 525 µm. A 1-µm thick layer of thermally grown SiO$_2$ was deposited on the polished side of the silicon. Beneath the wafer, a 1-mm thick glass layer was placed to limit the current, followed by a copper ground-side electrode.

\begin{figure}
    \centering
    \begin{tikzpicture}
    
        \node at (0,0) {\includegraphics[width=8cm]{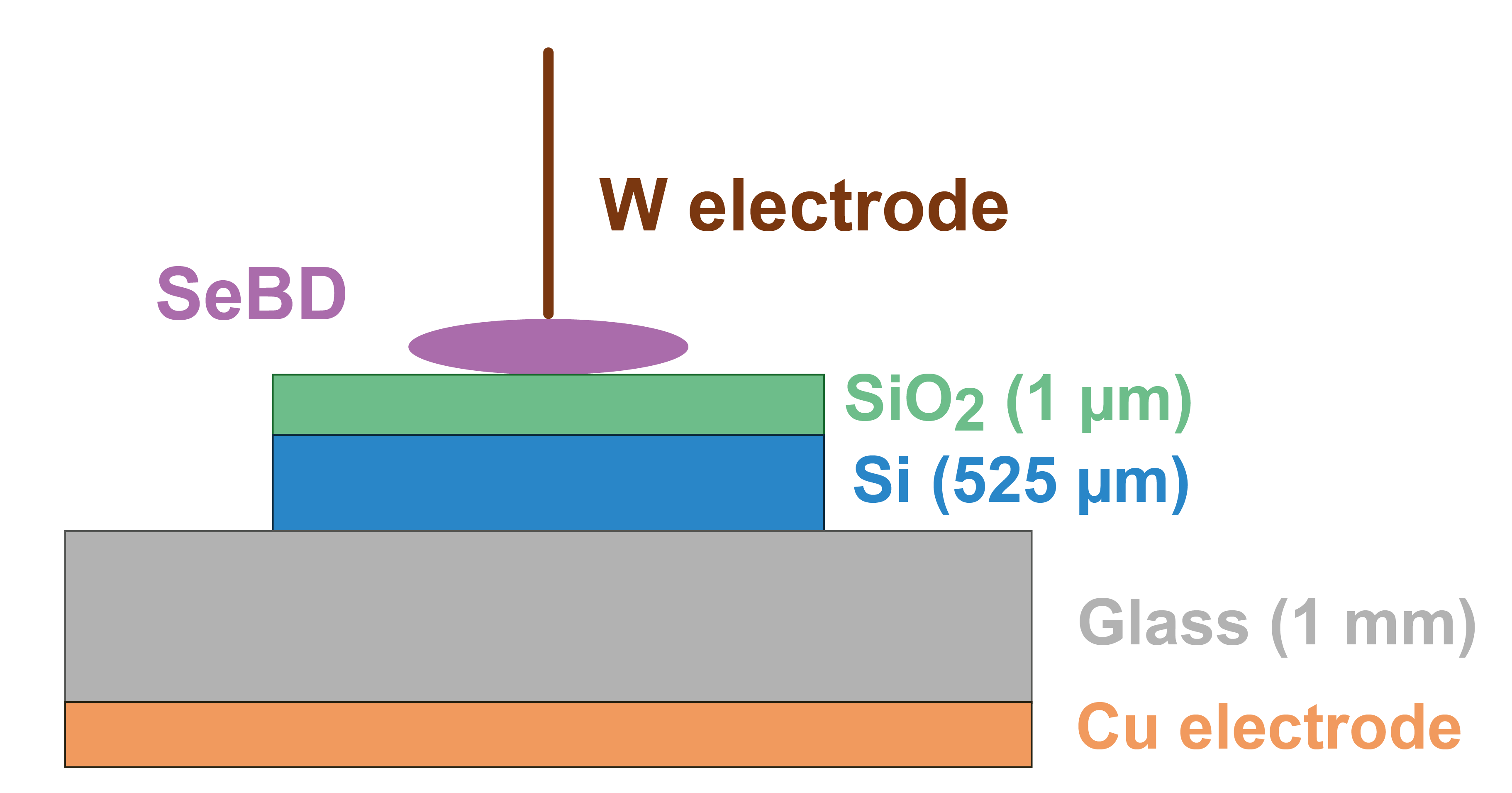}};
        
        \node at (0,-6.5) {\includegraphics[width=12cm] {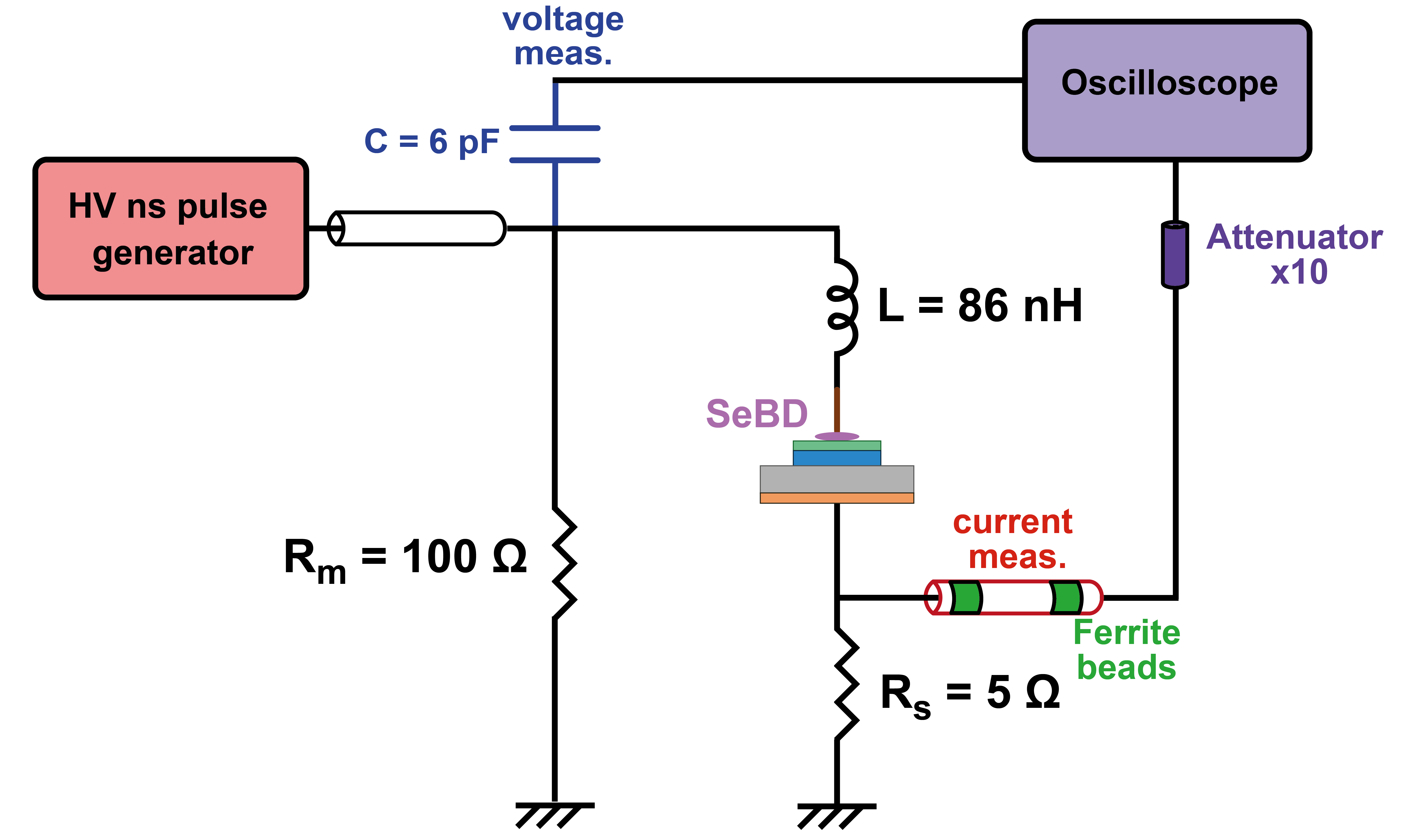}};
        
    \end{tikzpicture}
    \caption{\emph{Top:} Reactor geometry of the surface SeBD. 
    \emph{Bottom:} 
    Schematic diagram of the surface SeBD electrical circuit. }
    \label{2elecdiag}
\end{figure}

High voltage was applied to the pin electrode using a generator (FID Technology), transmitted first through a coaxial cable and then a wire segment whose inductance will be specified later, as shown in Figure \ref{2elecdiag}. The generator delivered 20-ns pulses with 5-ns rise time, 8-ns fall time and +2 kV amplitude at a repetition frequency of 100 Hz (Figure \ref{3curve}). A resistance R$_m$ = 100 $\Omega$ (Caddock MP9100) in parallel with the discharge circuit served to match the impedance of the coaxial cable from the generator. The applied  voltage at the end of the coaxial cable was measured using a passive probe (LeCroy PPE6KV, 400 MHz bandwidth, 6 pF capacitance). Additionally, a sensing resistance R$_s$ = 5 $\Omega$ (Caddock MP725) was inserted between the copper plate and the ground to measure the current through the reactor using a coaxial cable (RG400). The measured signal was attenuated by a factor of 10. Ferrite beads were placed around this cable to attenuate high-frequency electromagnetic noise (Fair-Rite 31 material, filtering frequencies $1-300$ MHz). High accuracy of the electrical measurements was ensured by replacing the SeBD with resistive, inductive, and capacitive load impedances. The measured current $i$ and voltage $V$ waveforms were then verified to be consistent with the expected circuit response. The inductive load was formed by short-circuiting the pin to the copper plate, yielding an inductance of the wire electrode of $L = 86$ nH while confirming that $V = L \frac{di}{dt}$ with high fidelity as in \cite{orriere2018confinement}.

\subsection{Fast imaging and Optical Emission Spectroscopy} \label{iccdoes}

Second, the optical diagnostics bench combined fast imaging and OES into a single system. The plasma reactor was placed at the focal plane of a $15\times$ magnification UV-reflective microscope objective (Beck Optronic 5002) with a working distance of 23.2 mm. The collimated plasma emission was focused using a UV-achromatic doublet lens L3 (Newport, f = 20 cm) and guided by UV-enhanced mirrors M1 and M2 (Thorlabs) to the spectrograph (Princeton Instruments, SpectraPro HRS-500) equipped with an intensified charge-coupled device (ICCD) camera (Princeton Instruments, PI-MAX 4). To perform OES measurements, a movable entrance slit of the spectrometer was inserted, and its width was set to 20 µm. A diffraction grating ruled with 1200 grooves/mm and blazed at 500 nm was selected to provide a suitable spectral range for the emission features of interest. This system thus allowed the imaging of both the plasma and its emission spectrum, with a maximum temporal resolution of 400 ps.

\begin{figure}[!t]
\centering
\includegraphics[width=16.5cm]{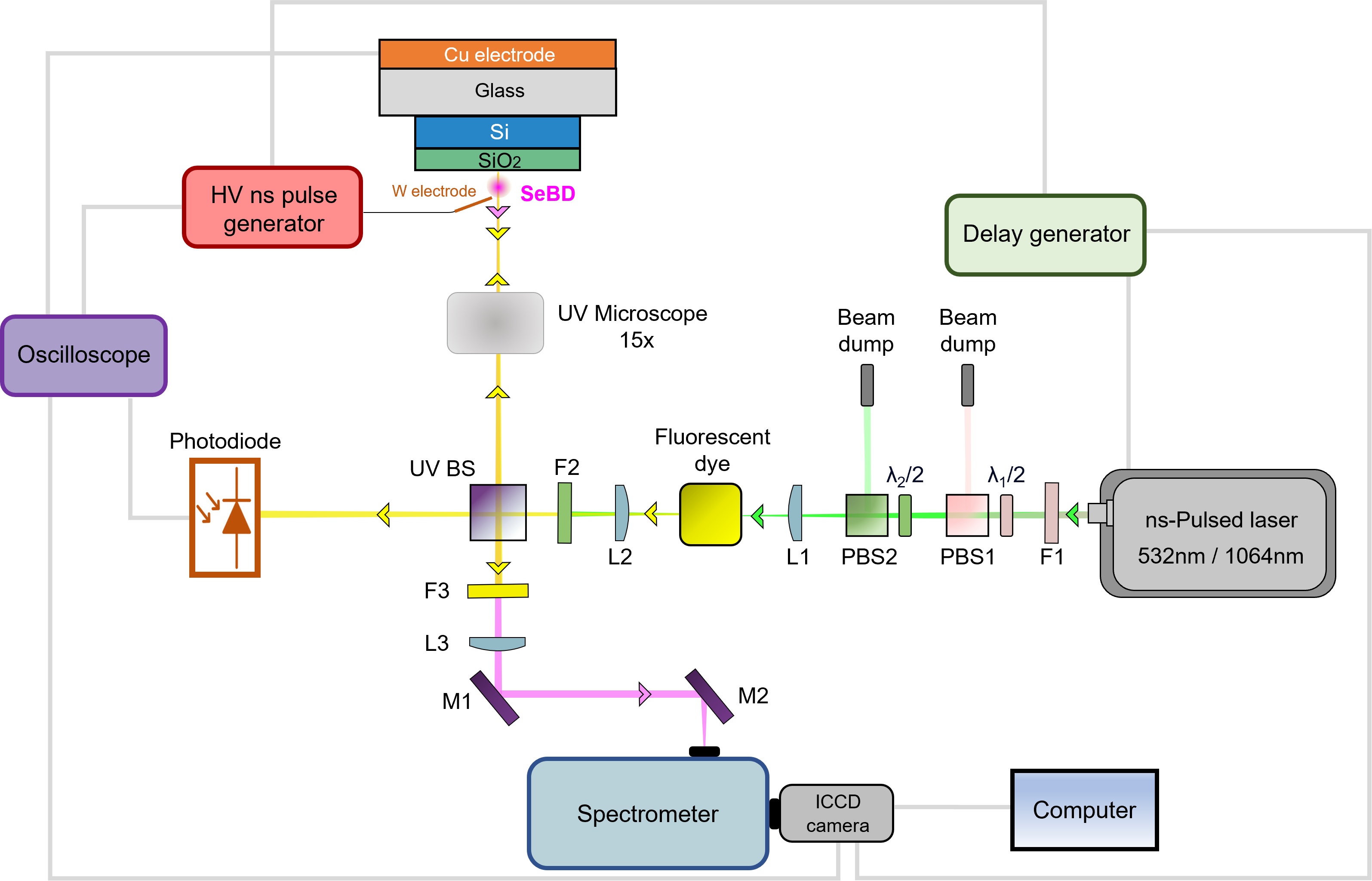} 

\caption{ Schematic diagram of the experimental setup illustrating the systems for plasma generation, fast imaging/OES, and external light illumination. The synchronization is illustrated by the connections between devices.}
\label{3setup}
\end{figure}

For OES, we analyzed the first negative system (FNS) of ionized nitrogen, specifically the \( N_2^+ (B^2\Sigma_u^+, v'=0 \rightarrow X^2\Sigma_g^+, v''=0) \) transition at 391.4 nm, and the second positive system (SPS) of neutral nitrogen, specifically the \( N_2 (C^3\Pi_u, v'=1 \rightarrow B^3\Pi_g, v''=4) \) transition at 399.8 nm to obtain qualitative information on the local electric field. In nanosecond pulsed discharges, these excited states are populated initially by electron impact excitation and direct ionization from the ground state of nitrogen molecules :
\[
\mathrm{N}_2 (X^1\Sigma_g^+) + e^- \rightarrow \mathrm{N}_2 (C^3\Pi_u) + e^-
\]
\[
\mathrm{N}_2 (X^1\Sigma_g^+) + e^- \rightarrow \mathrm{N}_2^+ (B^2\Sigma_u^+) + 2e^-
\]

Creating the $N_2^+ (B)$ state requires higher electron energies ($\sim 18.8$ eV) compared to the $N_2 (C)$ state ($\sim 11$ eV), making the emission intensity ratio FNS/SPS from these two transitions a useful indicator of the electron energy distribution and thus the local reduced electric field \cite{bruggeman2013atmospheric}.   For highly non-uniform and transient discharges, the method is subject to a number of requirements and considerations. These include the relationship between the electric field and the electron energy distribution function, the kinetic model of deexcitation of these states, the locality of the plasma emission and electric field within the limits of the spatial resolution, and whether stepwise ionization and the time derivatives of the emission intensities need to be taken into account \cite{paris2004measurement, lepikhin2022electric, goldberg2022electric, brisset2019modification}. The required conditions have generally been met in nanosecond discharges. However, the proximity to surfaces introduces cathode sheath dynamics and charge transfer which may impact the local field approximation \cite{jansky2021electric}. Furthermore, fast plasma dynamics and collisional quenching at atmospheric pressure imply that emission may not instantaneously reflect only the local excitation. In this context, the perfect uniformity of the SeBD mitigates problems related to spatial resolution.
The use of a short integration time of 400 ps brings the measured FNS/SPS into closer correspondence with the value of the electric field. Despite the favourable conditions of this study, the aforementioned problems likely still introduce enough inaccuracy to warrant caution when using FNS/SPS to derive the absolute value of the reduced electric field, particularly without the help of an accompanying collisional-radiative model. However, relative changes to the electric field are less sensitive to these factors,
especially since we are only concerned with how the reduced electric field responds to a perturbation of the SeBD by irradiation. The kinetics of quenching and surface-related processes likely remain fundamentally the same with and without irradiation. Therefore we will use the intensity ratio FNS/SPS in this comparative way.

\subsection{External light illumination} \label{illumination}

Third, to illuminate the wafer, a diode pumped solide-state laser (ElforLight SPOT-10-200-532) with a pulse duration of $2-3$ ns at 532 nm and $4-5$ ns at 1064 nm and repetition frequency of 100 Hz was used to simulate the photonic effect between the plasma and the silicon. At the laser output, the collimating lens of the laser and its optical path length were modified over the course of the experiments (not shown in Figure \ref{3setup}), which could have introduced variation of the beam divergence upon incidence on the SeBD, although the laser spot size was held constant. The impact of this variation will be reflected in the uncertainty assigned to the relevant experimental data. Both beams were produced simultaneously, but only the 532 nm or 1064 nm beam continued on the optical path upon inserting filter F1 (Thorlabs), which was either a reflective band pass filter at 532 nm or reflective notch filter at 532 nm, respectively.

To investigate the influence of varying photon flux in this study, two polarizing beamsplitting cubes (PBS1 and PBS2 for the 1064 nm and 532 nm laser beams, respectively) combined with half-wave plates ($\lambda_1/2$ and $\lambda_2/2$ for the 1064 nm and 532 nm laser beams, respectively) were placed at the laser output to adjust the laser power. The remaining portion of each beam was directed to a beam dump. When necessary, optical density filters (Thorlabs, not shown in Figure \ref{3setup}) were inserted after the polarizing beamsplitters to further reduce the laser power.

Additionally, we explored the effect of different photon energies exceeding the silicon bandgap energy $E_g$ = 1.12 eV to ensure the excitation of electron-hole pairs. For this, we selected initially between the available 532 nm and 1064 nm laser beams by using filter F1, as mentioned previously. To illuminate at additional wavelengths, we focused the 532 nm beam using lens L1 (Thorlabs, f = 6 cm) into a cuvette containing fluorescent dye solution in ethanol. The optimal dye concentration was determined by analyzing the fluorescence spectra recorded with the spectrometer, selecting for the maximum energy conversion from the 532 nm pump beam. The chosen dyes were fluorescein (Sigma-Aldrich F6377) emitting at $540 - 620$ nm and Nile Red (MP Biomedicals) emitting at $600 - 750$ nm. The resultant emission was filtered (filter F2), first by the 532 nm notch filter to suppress residual pump beam energy. Second, for fluorescein, a long-pass absorptive filter (Schott OG570) was added to create spectral separation from the case of 532 nm illumination by suppressing fluorescence in the $540 - 560$ nm range (Figure \ref{3f(w)}).

\begin{figure}[!htbp]
\centering
\includegraphics[width=12cm]{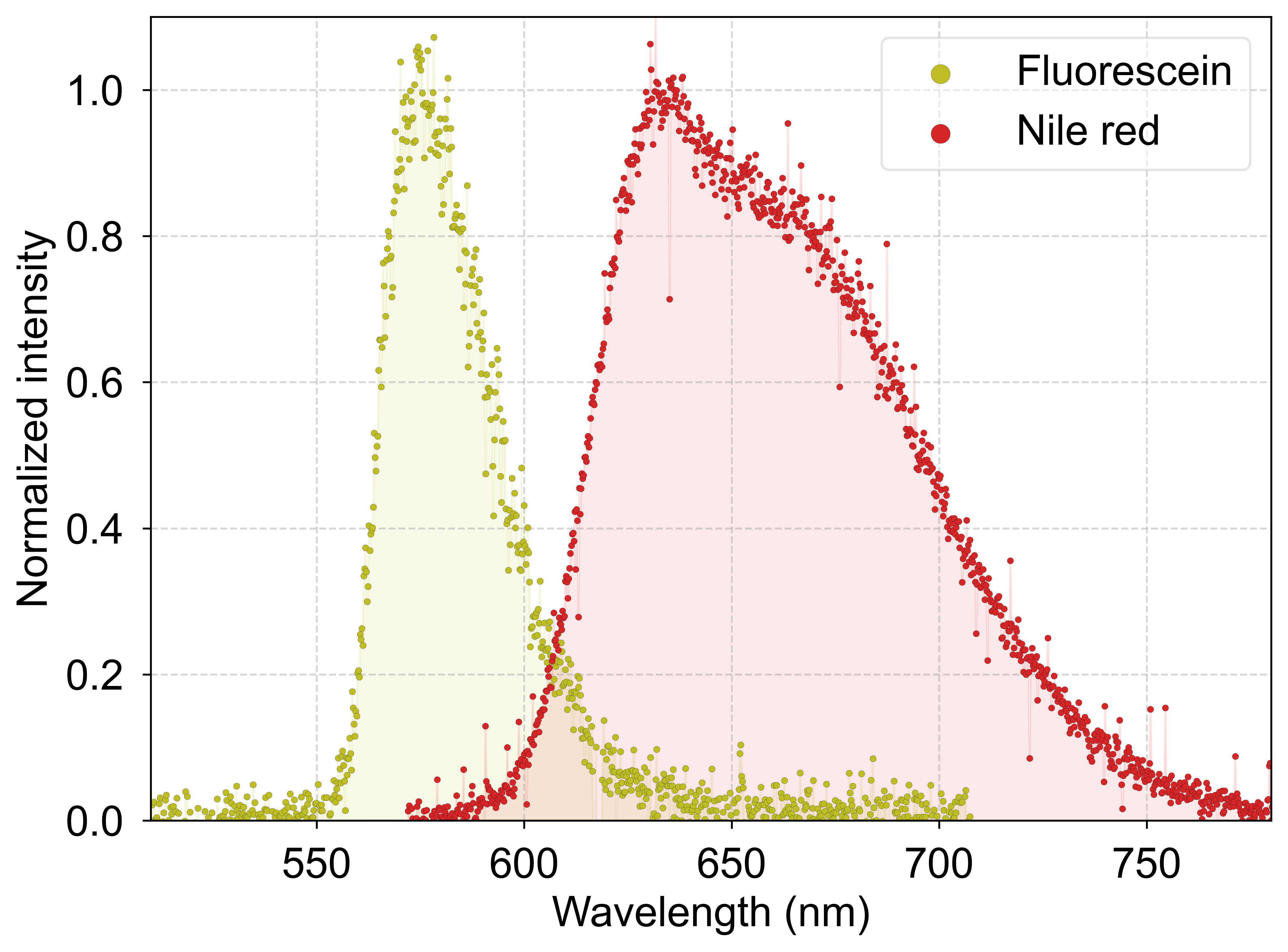} 

\caption{Fluorescence spectra of fluorescein and Nile Red dyes after passing through Filter 2, recorded with the spectrometer using a 10 ns exposure time to capture the emission near the beginning of the fluorescence pulse of each dye, as shown in Figure \ref{3f(t)}.} 
\label{3f(w)}
\end{figure}

After passing through the dye solution, lens L2 (Thorlabs, f = 12.5 cm) was employed to collimate the emerging beam. A non-polarizing UV-beamsplitter $50:50$ (UV-BS) directed half of the beam onto the wafer via the microscope objective previously mentioned, while the other half was sent to a photodiode (Thorlabs DET025A/M silicon PIN photodetector, 2 GHz bandwidth) positioned behind the beamsplitter to measure the beam power. To determine the photon flux reaching the wafer surface, we calibrated the photodiode against a power meter (Thorlabs S121C silicon photodiode power sensor). Because the power meter was inaccurate at low laser repetition frequencies (100 Hz in our case), this involved measuring the average laser power at higher frequencies (above 5 kHz) directly at the wafer position. Simultaneously, we recorded the photodiode signal and divided the time-integrated signal by the energy per pulse determined using the power meter to find the conversion factor of calibration. At 100 Hz, we used only the photodiode and applied this factor to calculate the power reaching the wafer. A typical value for this factor lies in the range of $0.09-0.29$ V/W, with a relative uncertainty between 2\% and 14\%. This uncertainty will be accounted for in the final results.

Finally, to block any laser light or fluorescence reflected off the Si-SiO$_2$ interface, appropriate optical filters (Filter 3) were placed in front of the spectrometer, ensuring that only plasma emission was captured by the ICCD camera. Filter 3 consisted of either the 532 nm notch filter (when using this wavelength as external irradiation), an absorptive band pass filter (Schott UG1 or BG12 for fluorescein or Nile Red emission, respectively) or an absorptive short pass filter (Schott KG3 when using 1064 nm as external irradiation). 

Both fluorescence emissions exhibit broader spectra and are less coherent than the laser beams (Figure \ref{3f(w)}). Photodiode measurements also showed that the laser pulse duration is slightly longer at 1064 nm than at 532 nm (Figure \ref{3f(t)}). Also, fluorescence lasts longer than the laser pulses : more than 20 ns for fluorescein and 10 ns for Nile Red.

\begin{figure}[!htbp]
\centering
\includegraphics[width=12cm]{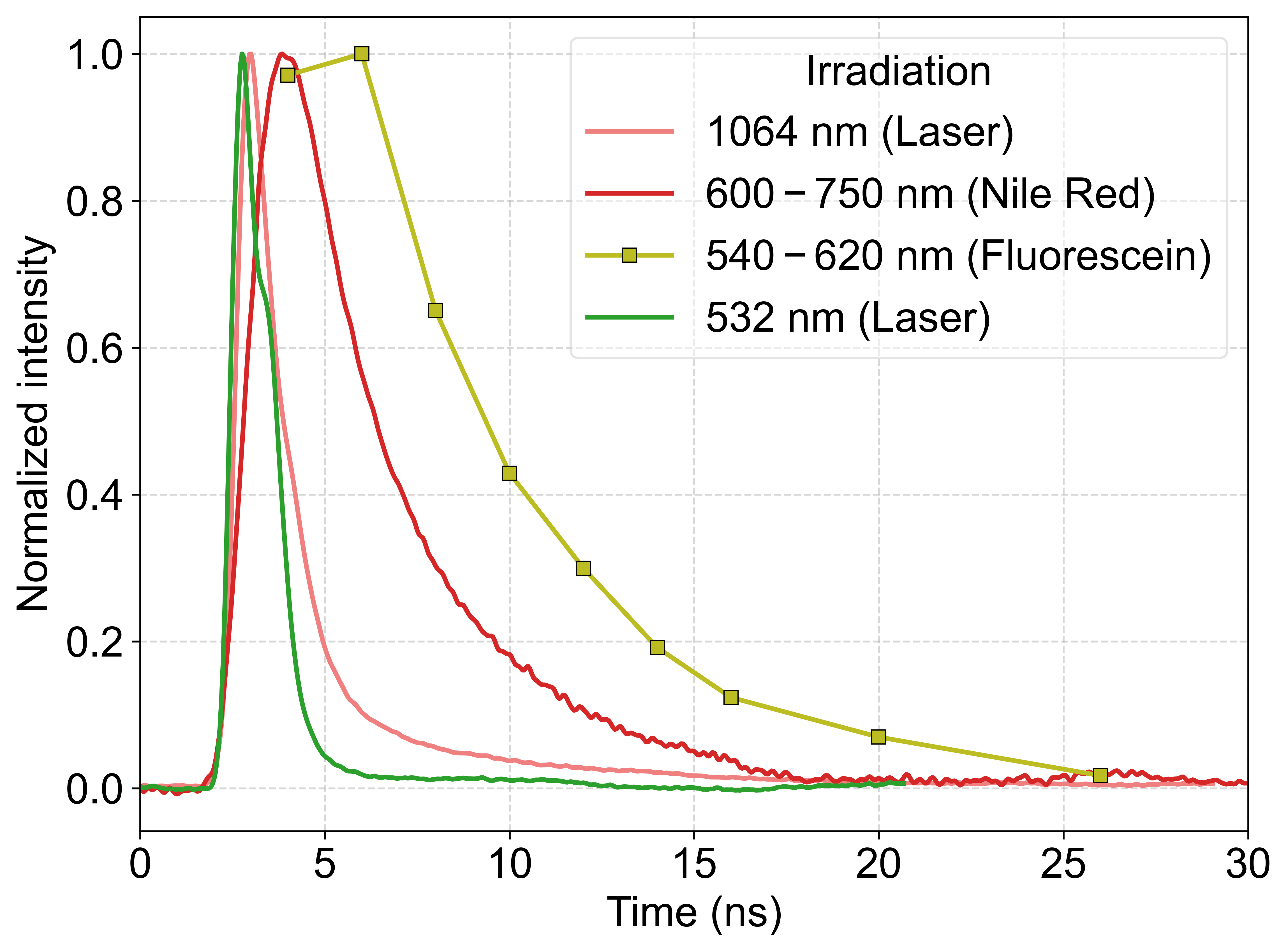} 

\caption{Time evolution of different sources of external irradiation. All signals were acquired with the photodiode, except for fluorescein, which was recorded without F2 using the spectrometer with an exposure time of 10 ns. } 
\label{3f(t)}
\end{figure}

The fluences used in our study covered the range from $10^{-9}$ to $2 \times 10^{-3}$ J/cm$^2$. At 532 nm with nanosecond pulses, fluences in this range lie far below the typical melting/ablation thresholds for crystalline silicon, which are on the order of $1-100$ J/cm$^2$ \cite{beranek2023melting}. Consequently, single-pulse exposure in our range produces predominantly photo-excitation and negligible bulk heating \cite{caruso2022ultrafast}.
Cumulative pulses can gradually lower thresholds \cite{fu2010experimental,vladoiu2009influence}, but reaching morphological modification usually requires much higher per-pulse fluence \cite{chen2022nanosecond,binetti2016picosecond}.

All signals, including the high-voltage pulse, current, camera gate monitor and photodiode were obtained with a 2 GHz bandwidth oscilloscope (Rhode \& Schwartz RTO2024). A delay generator (SRS DG645) synchronized the high-voltage pulse generator, ICCD camera and laser at a frequency of 100 Hz.

\section{Methods and Results}

In this section, we will first describe the characteristics of the SeBD obtained under the defined experimental conditions without irradiation (Section \ref{woirradiation}). This will facilitate the presentation of the methods of data treatment used to quantify plasma optical emission (Section \ref{radialprofile}) and irradiation timing and positioning (Section \ref{timeposirrad}). Then, we will analyze the plasma emission intensity (Section \ref{irradiation}), electric field (Section \ref{efield}), and discharge energy (Section \ref{energy}) under external irradiation synchronized with the discharge, which excites electron-hole pairs in silicon. The analysis will primarily focus on the effects of irradiation fluence and wavelength.

\subsection{Discharge characteristics} \label{discharge}

\subsubsection{SeBD without irradiation} \label{woirradiation}

Figure \ref{3curve} \emph{Top} presents the current-voltage waveforms under the unperturbed discharge conditions (without external illumination) used throughout this study. Figure \ref{3curve} \emph{Bottom} provides time-resolved top-view images of plasma propagation.  Pulses of $1.95 \pm 0.05$ kV amplitude were applied at 100 Hz repetition frequency.
The current initially increases due to capacitive coupling before breakdown occurs at $t_0 = 1.9$ ns. After breakdown, a distinct two-phase behaviour is observed. First, the positive discharge ignites at the breakdown voltage $V(t_0)$ = 1.4 kV with a filled circular structure. The current subsequently reaches its maximum value of 2.5 A at $t$ = 3.2 ns. The plasma then transitions into a propagating ring-shaped ionization wave at $t$ = 4.3 ns with $100 \pm 13$ µm thickness. It starts expanding at a speed of $1.25 \times 10^5$ m/s and then slows progressively to $5 \times 10^4$ m/s. This positive discharge lasts for 13.2 ns and achieves a maximum extension of $500 \pm 13$ µm. In addition to the ionization wave, a "corona" surrounds the electrode. During this phase, the current waveform oscillates due to the SeBD circuit shown in Figure \ref{2elecdiag} but not due to any parasitic circuit elements, which have been eliminated through the circuit test procedure described in Section \ref{sec:plasmagen}. These oscillations may result from the coupling between the inductance of the wire electrode and a capacitance created by the plasma. In the subsequent phase, the plasma reignites as a negative discharge at $t$ = 15.1 ns with a negative current peak of $-1.2$ A , following a propagation pattern similar to the first phase but with a more diffuse appearance. The negative ring shape appears at $t$ = 25.9 ns with a thickness of $120 \pm 13$ µm. This phase lasts for more than 25 ns. The overall discharge behaviour is similar to that reported in \cite{darny2020uniform}, despite differences in pulse rise time and duration, which were around 10 ns and 15 ns respectively.

The electrical energy of the discharge shown in Figure \ref{3curve} was measured to be $8.36 \pm 0.2$ µJ. More generally, the discharge energy typically lies within the range of $9 \pm 0.9$ µJ. The uncertainty of $\pm$0.2 µJ corresponds to the standard deviation of the measurement itself, encompassing both instrumental noise and the intrinsic variability of the plasma, which is most pronounced after the second breakdown at $t$ = 15.1 ns. The remaining uncertainty up to $\pm$0.9 µJ is primarily attributed to variations in the experimental conditions, in particular the mechanical contact and positioning of the electrode on the wafer, as well as surface purity, cleanliness, and prior plasma exposure at a given location on the wafer.

\begin{figure} [!htbp]
    \centering
    \begin{tikzpicture}
        \node at (0,0) {\includegraphics[width=15cm]{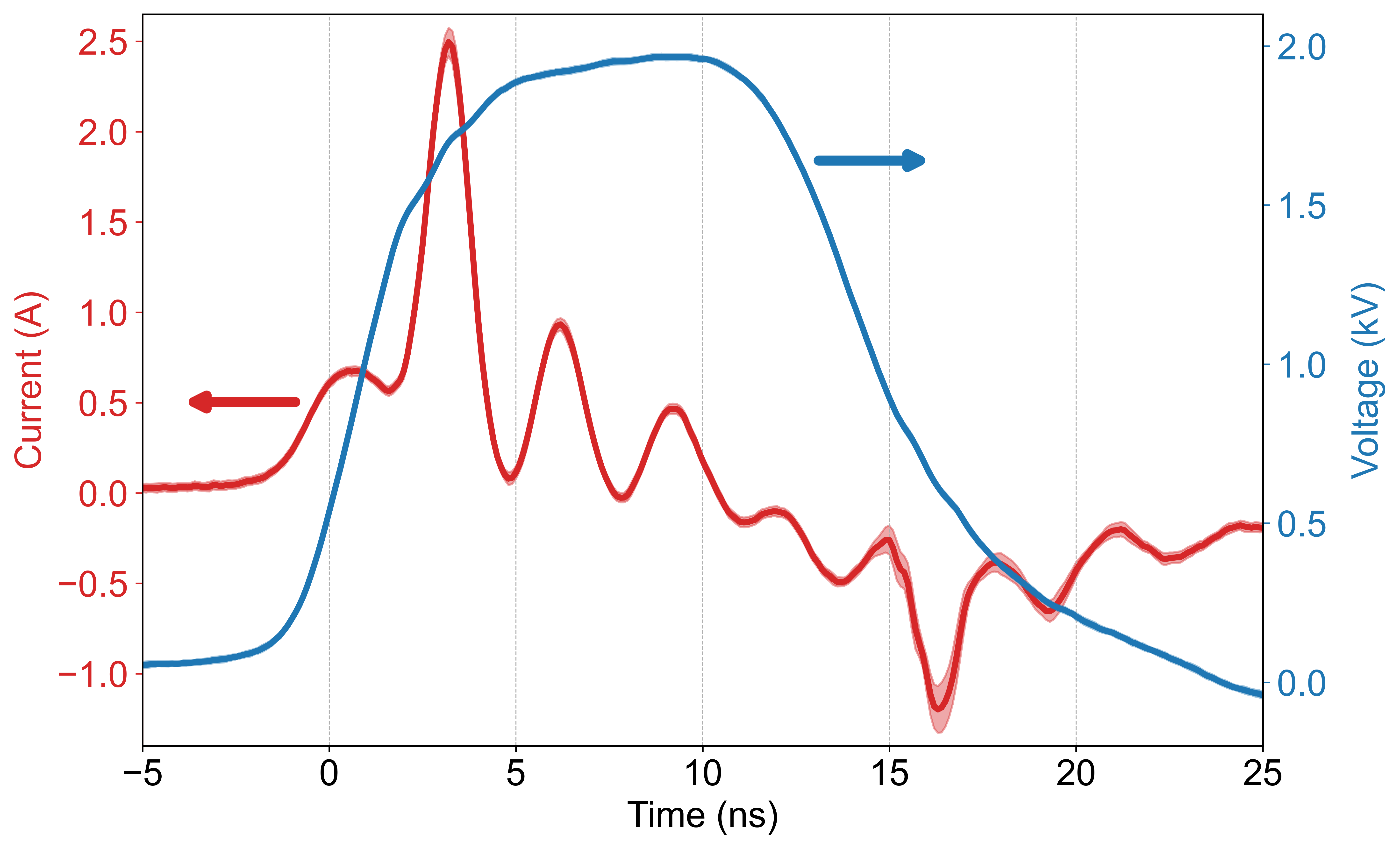}};

        \node at (0,-10) {\includegraphics[width=15cm]{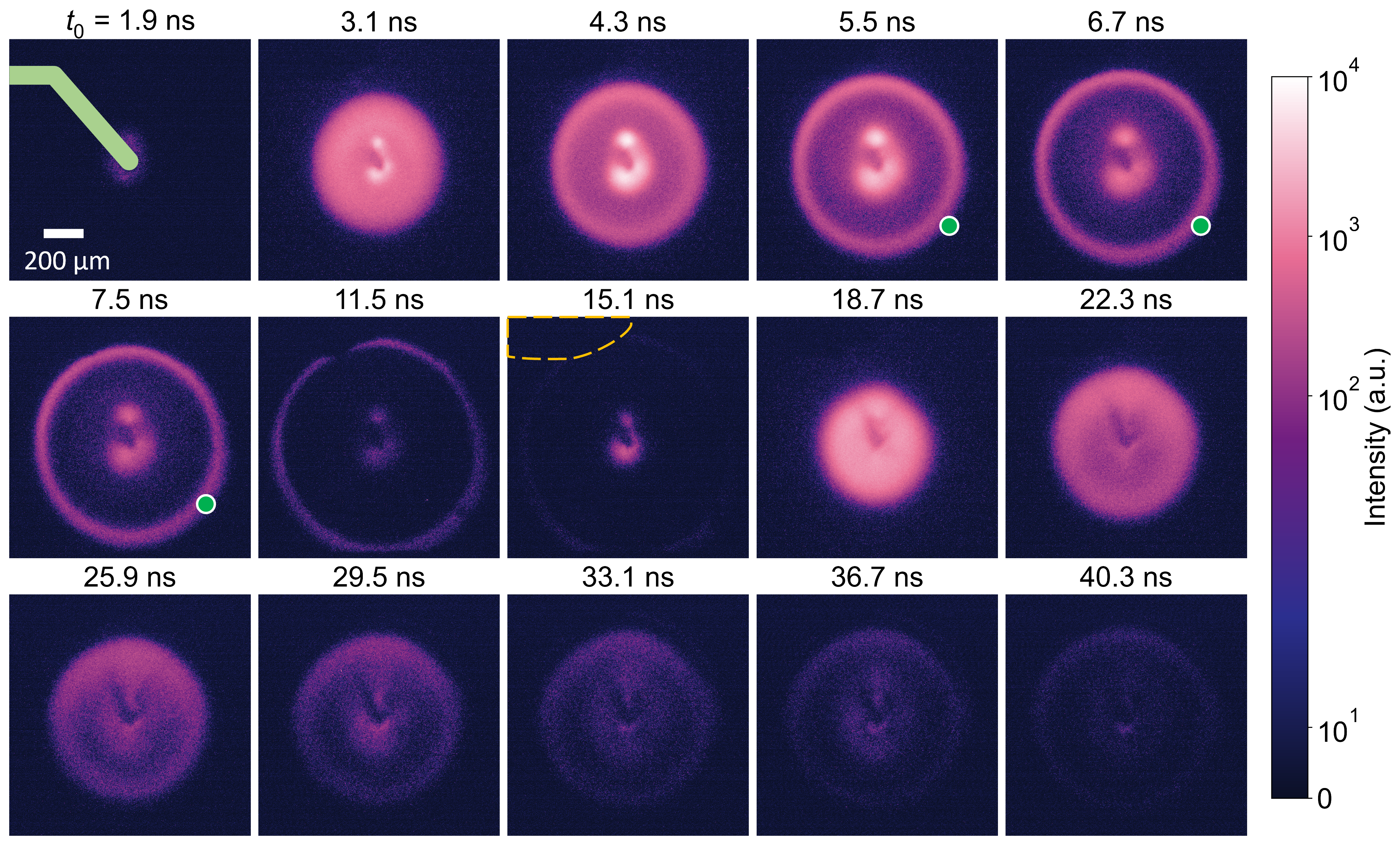}};

    \end{tikzpicture}
    
    \caption{\emph{Top :} Current-voltage waveforms of the discharge averaged over 100 acquisitions. \emph{Bottom :} Time-resolved images of discharge propagation. Individual frames were obtained using an exposure time of 400 ps and accumulated on the CCD over 300 discharge events. The colorbar shows the scale of relative intensity. The shadow of the electrode is represented in the first image at 1.9 ns. The 532 nm laser spot initiates at 5.3 ns (not shown), indicated by the small circle near the bottom-right corner of the images at 5.5 ns, 6.7 ns and 7.5 ns. Plasma emission is partially blocked from view by an optical obstacle at 11.5 ns and 15.1 ns (dashed outline).}
\label{3curve} 
\end{figure}

\subsubsection{Radial profile of plasma optical emission} \label{radialprofile}

To quantify the plasma emission intensity, we first determined the center of the circle using a RANSAC-based \cite{fischler1981random} circle fitting algorithm applied to a binary thresholded version of the image. A circular geometry of the plasma was assumed during this fitting process. A specific angular sector spanning the laser-exposed region was then selected. Integration of the signal over the angular coordinate yielded the radial profile. The radial intensity profile within this sector was smoothed using a one-dimensional uniform filter implemented in Python \cite{virtanen2020scipy}. This filter performs a local averaging of neighboring points within a sliding window, effectively reducing high-frequency noise while preserving the overall shape of the emission profile. The resulting signal was finally baseline-corrected using the adaptive airPLS algorithm \cite{zhang2010baseline}. An example profile is shown in Figure \ref{3arc}. From 0 to 200 µm, the emission originates from the plasma corona. From 340 to 540 µm, the emission corresponds to the annular ring exposed to the laser. The ring section of the processed profile was then fitted using a single Gaussian function, whose center position $r_c$ and standard deviation $\sigma$ define the radial position and thickness of the ring, respectively, as shown in Figure \ref{3arc}. The integrated intensity was computed in the radial coordinate within $\pm 3\sigma$, ensuring the intensity falls to zero on either side of the ring. However, in some cases the plasma emission extends beyond $+3\sigma$ (as described in Section \ref{intensity}). In such instances, this additional contribution was also included.

\begin{figure}[!htbp]
    \centering
    \begin{tikzpicture}
        \node at (0,0) {\includegraphics[width=7.5cm]{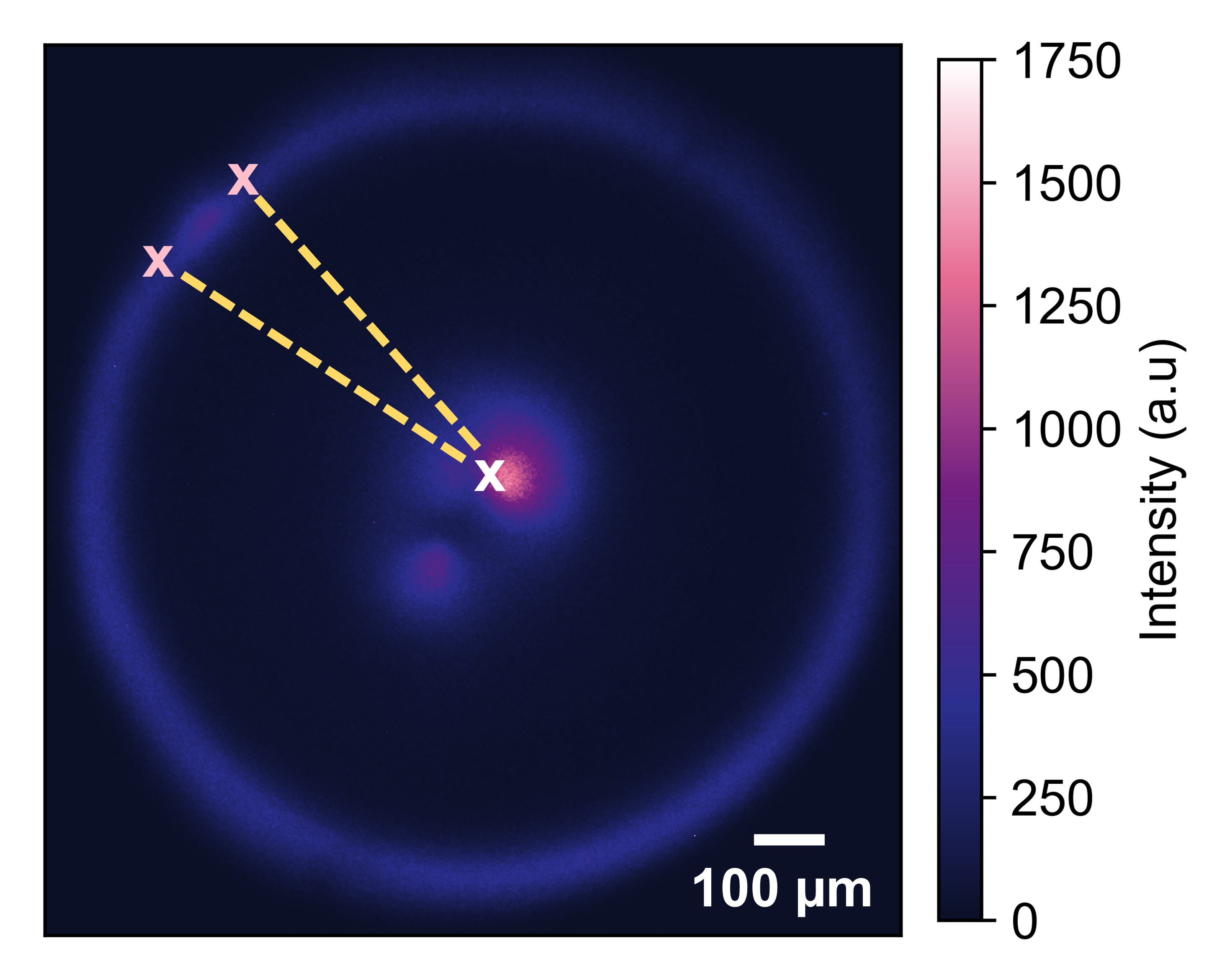}};

        \node at (8,0) {\includegraphics[width=8.5cm]{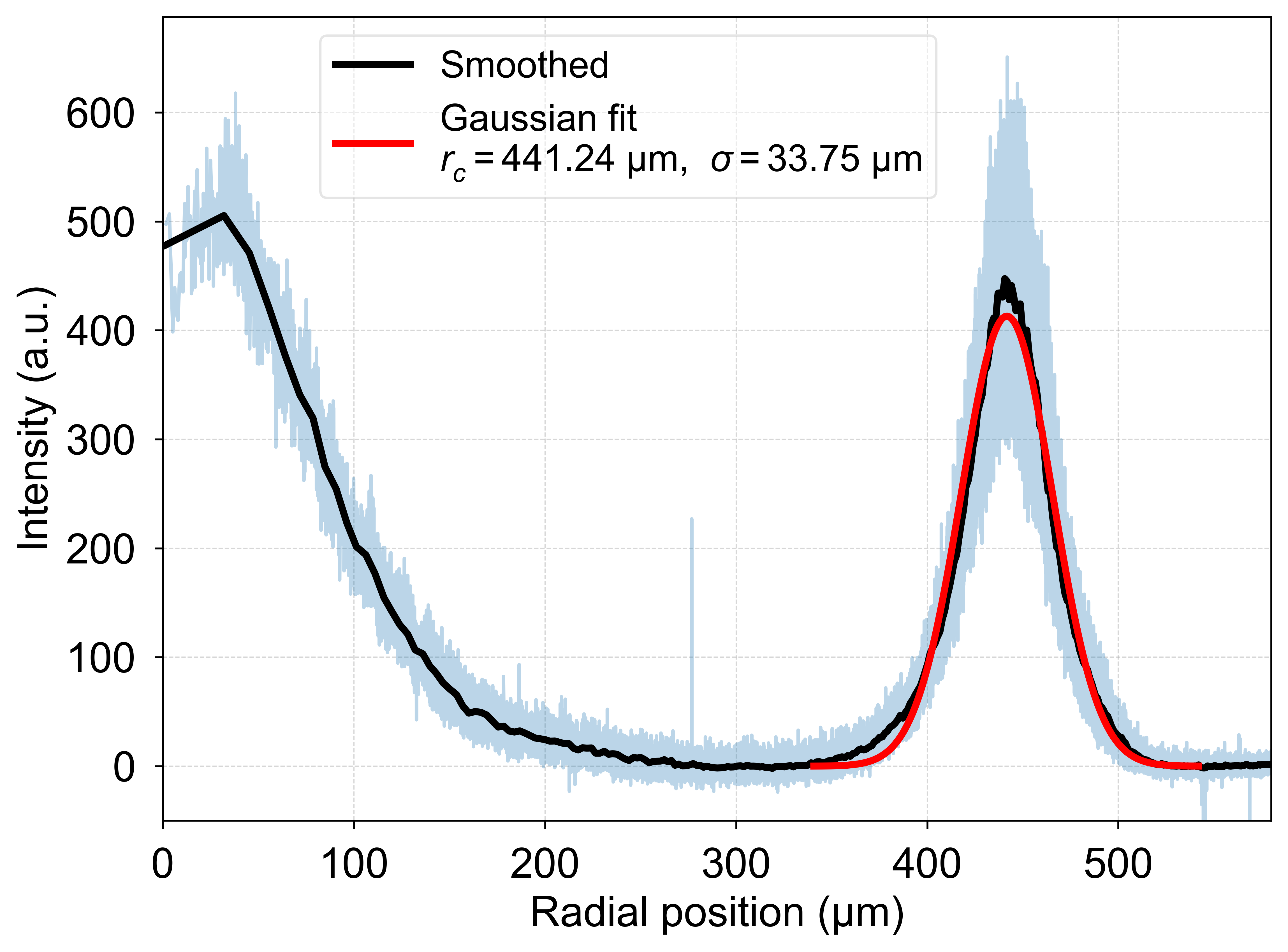}};

    \end{tikzpicture}
    
    \caption{\emph{Left :} Example image of the SeBD under external irradiation of fluence $F$ = 2.33 µJ/cm$^2$ per pulse at 532 nm, along with the angular sector (dashed) used to construct the radial profile of plasma emission intensity. The white cross at the center indicates the circle center, while the two pink crosses along the ring ionization wave front delimit the circular arc exposed to the laser. \emph{Right : } Radial profile of the plasma emission intensity. The Gaussian function (red) is fitted to the ring section of the smoothed data (black).}
\label{3arc} 
\end{figure}

\subsection{Emission intensity} \label{intensity}

\subsubsection{Time and position of irradiation} \label{timeposirrad}

The laser/fluorescence spot size is $S$ = $(5.03 \pm 1.26) \times 10^3$ µm$^2$ in area, as shown in Figures \ref{3curve} and \ref{2electrode}. The times $t_1$ and $t_2$ correspond to the onset of surface illumination and to the maximum spatial overlap between the ionization front and the irradiation spot, respectively. The location of the spot was chosen so that it appears in front of the ring-shaped ionization wave at $t_1$, in order to simulate the effect of plasma photons on the propagation (Figure \ref{2electrode}). 
For the 532 nm and 1064 nm laser beams as well as Nile Red fluorescence, the surface illumination starts at $t_1$ = 5.3 ns. However for fluorescein, this timing was adjusted to account for its longer fluorescence lifetime, as explained in Section \ref{illumination}. In this case, an additional delay of 2 ns was introduced, resulting in $t_1$ = 3.3 ns. For all wavelengths, $t_2$ was set to 7.3 ns so that most of the photon energy had been delivered to the surface by the time of maximum overlap. Consequently, $t_2 - t_1 = 2$ ns for the lasers and Nile Red cases, whereas $t_2 - t_1 = 4$ ns for fluorescein.

\begin{figure}[!htbp]
\centering
\includegraphics[width=7cm]{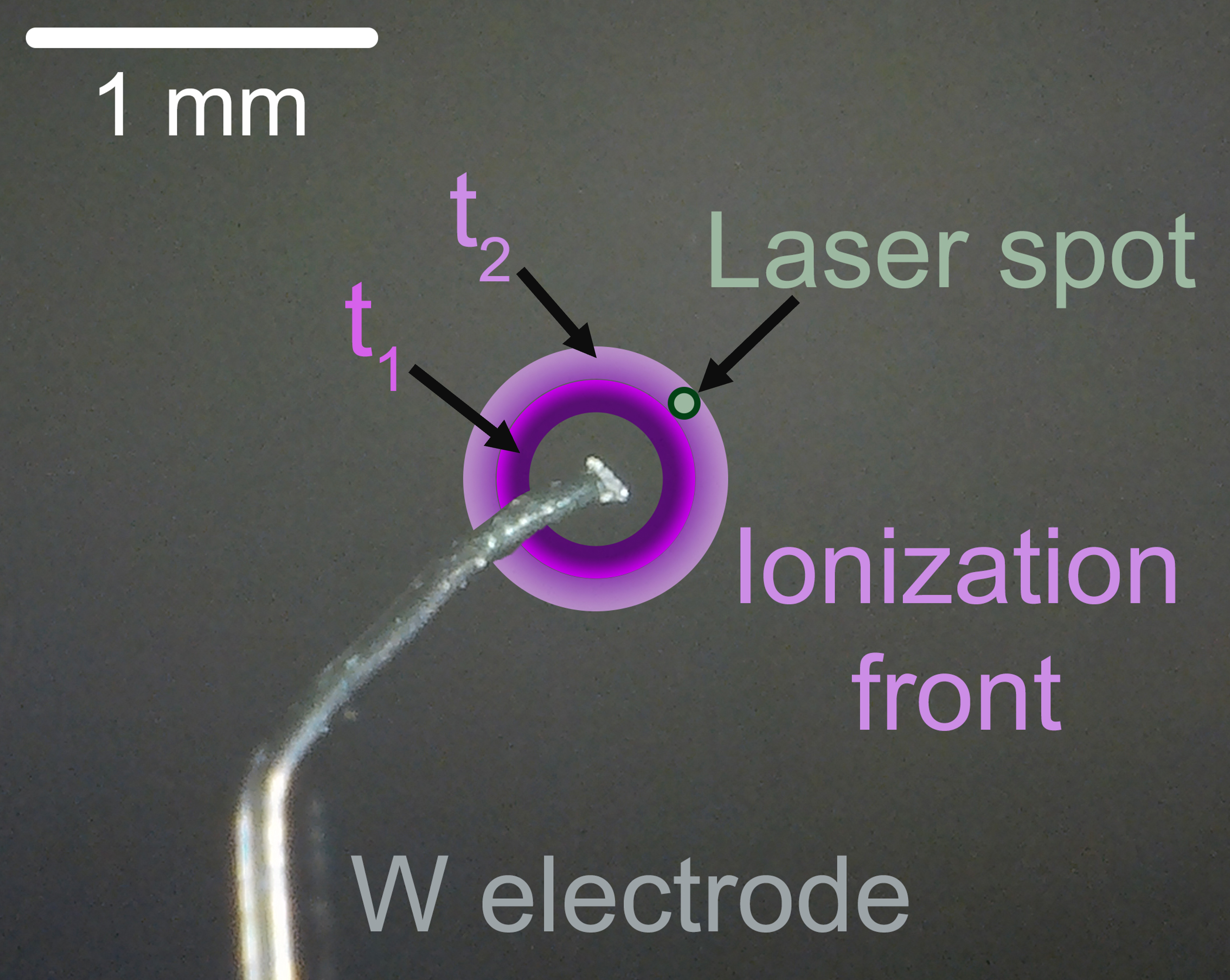} 

\caption{Top-view image of the tungsten electrode with its tip contacting the wafer, along with a schematic diagram of the ionization front at $t_1$ and $t_2$.} 
\label{2electrode}
\end{figure}

\subsubsection{Effect of irradiation} \label{irradiation}

Figure \ref{4laser} illustrates how plasma emission increases with illumination. Comparing panels (a), (b) and (c), we see that at irradiation below $F$ = 39.8 µJ/cm$^2$ at 1064 nm, the increase in plasma emission is localized to the region of the laser spot near the front of the ionization wave (pink arrows). The same behaviour is observed when using 532 nm below $F$ = 3.9 µJ/cm$^2$, comparing panels (e) and (f) (green arrows). This increase seems to come at the expense of the rest of the discharge which becomes slightly dimmer as a result. At a higher fluence, for example at $F$ = 130 µJ/cm$^2$ in panel (d) for irradiation at 1064 nm and at $F$ = 20.5 µJ/cm$^2$ in panel (g) for irradiation at 532 nm, the enhanced emission expands beyond the irradiation spot in both the radial and polar directions (dashed orange outline in panels (d) and (g)). 

Additionally, at higher laser fluences at 532 nm, our experiments revealed a memory effect not observed at lower fluences. As illustrated in panel (h) with the orange arrow, the same region of enhanced plasma emission shown in panel (g) experiences a reduction in intensity immediately after switching off the laser. This was also reported in \cite{darny2020uniform} but using a continuous wave laser at 532 nm rather than a pulsed laser. The plasma returns to its original form after several seconds to minutes.

\begin{figure}[!t]
\centering
\includegraphics[width=16.8cm]{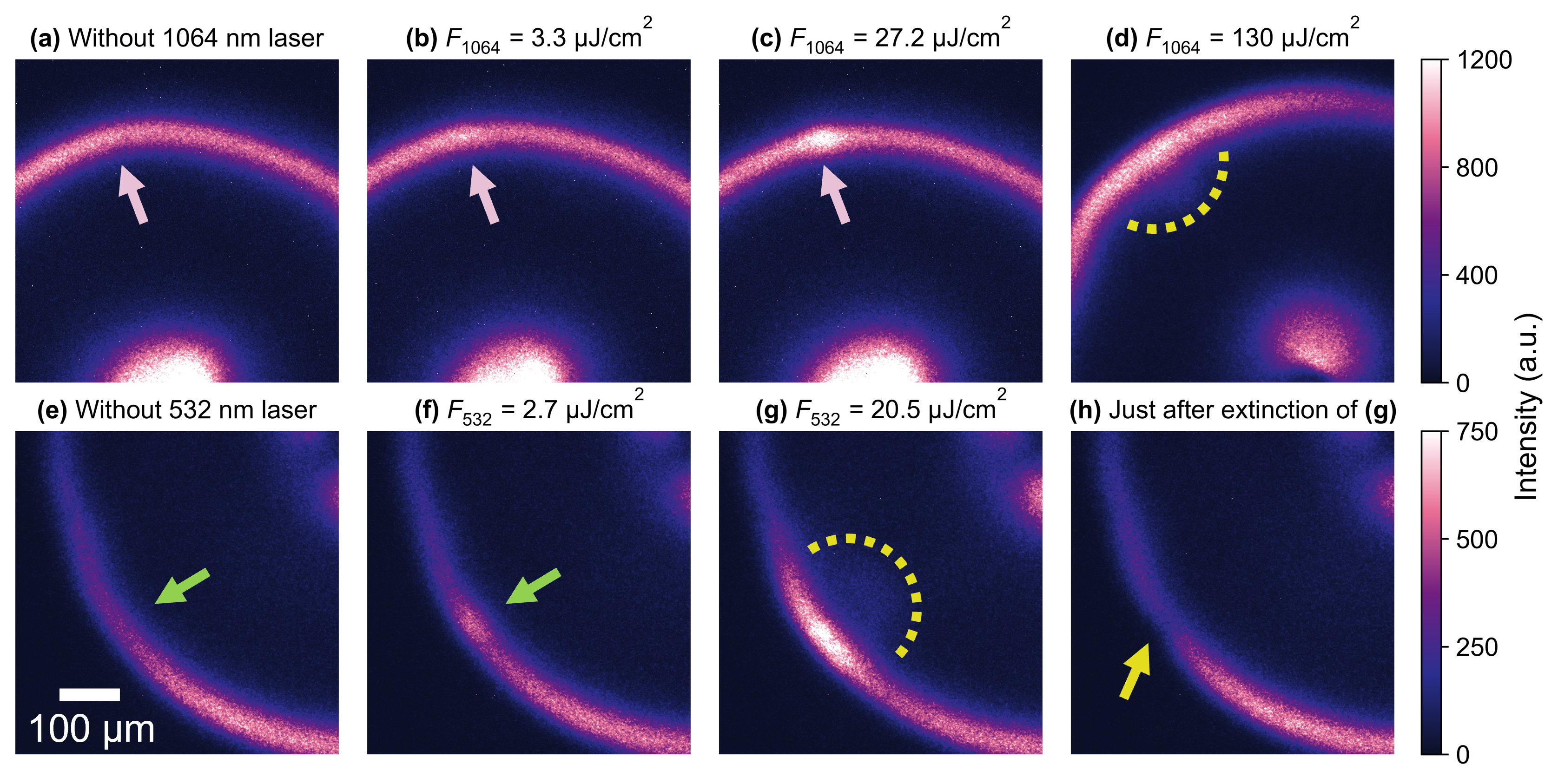} 

\caption{Images of the plasma acquired at $t$ = 7.3 ns with laser irradiation fluences $F$ at 1064 nm of (a) 0 µJ/cm$^2$, (b) 3.3 µJ/cm$^2$, and (c) 27.2 µJ/cm$^2$ and (d) 130 µJ/cm$^2$. The image in panel (d) was not acquired in the same experiment as (a), (b) and (c). Also shown are images obtained with 532 nm irradiation fluences $F$ of (e) 0 µJ/cm$^2$, (f) 2.7 µJ/cm$^2$, (g) 20.5 µJ/cm$^2$ as well as (h) immediately after stopping illumination at 20.5 µJ/cm$^2$. The arrows point to the laser spot position while the arrow in panel (h) points to the plasma location that is dimmer than in panel (e). The dashed orange outlines in panels (d) and (g) show the plasma emission increase expanding in the radial direction. Acquisitions were obtained using a 400 ps exposure time, accumulated 1000 times on-CCD and averaged 3 times.}
\label{4laser}
\end{figure}

Figure \ref{4emissionlarge} illustrates the relative increase in plasma emission $\frac{I - I_0}{I_0}$ as a function of the illumination fluence $F$ for various wavelengths ranging from 532 nm to 1064 nm. The plasma emission intensities $I$ and $I_0$ correspond to the cases with and without external illumination, respectively. The intensity $I_0$ corresponds to the unperturbed plasma conditions shown in Figure \ref{3curve}. 
No measurable increase in plasma emission is observed for any of the wavelengths below a threshold irradiation fluence. Plasma emission begins to increase at around $F$ = 3 µJ/cm$^2$ for 1064 nm, compared to approximately $F$ = 0.7 µJ/cm$^2$ for the other shorter wavelengths.

\begin{figure}[!htbp]
    \centering
    \begin{tikzpicture}

        \node at (0,0) {\includegraphics[width=13.5cm]{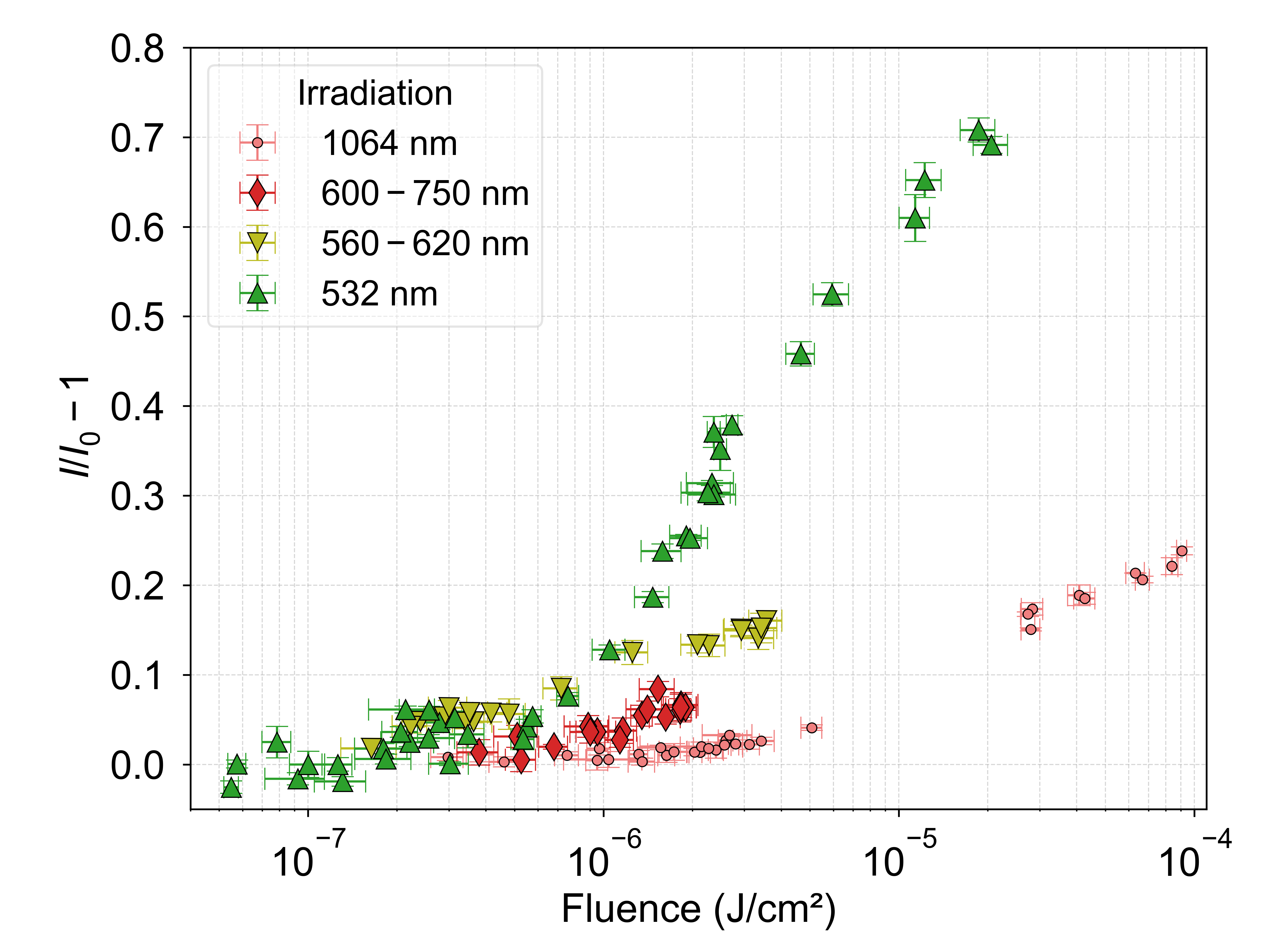}};

        \fill[violet!40, opacity=0.17]
            (-4.7,0.76) rectangle (1.15,-3.55);

        \node at (0,-10.2) {\includegraphics[width=13.5cm]{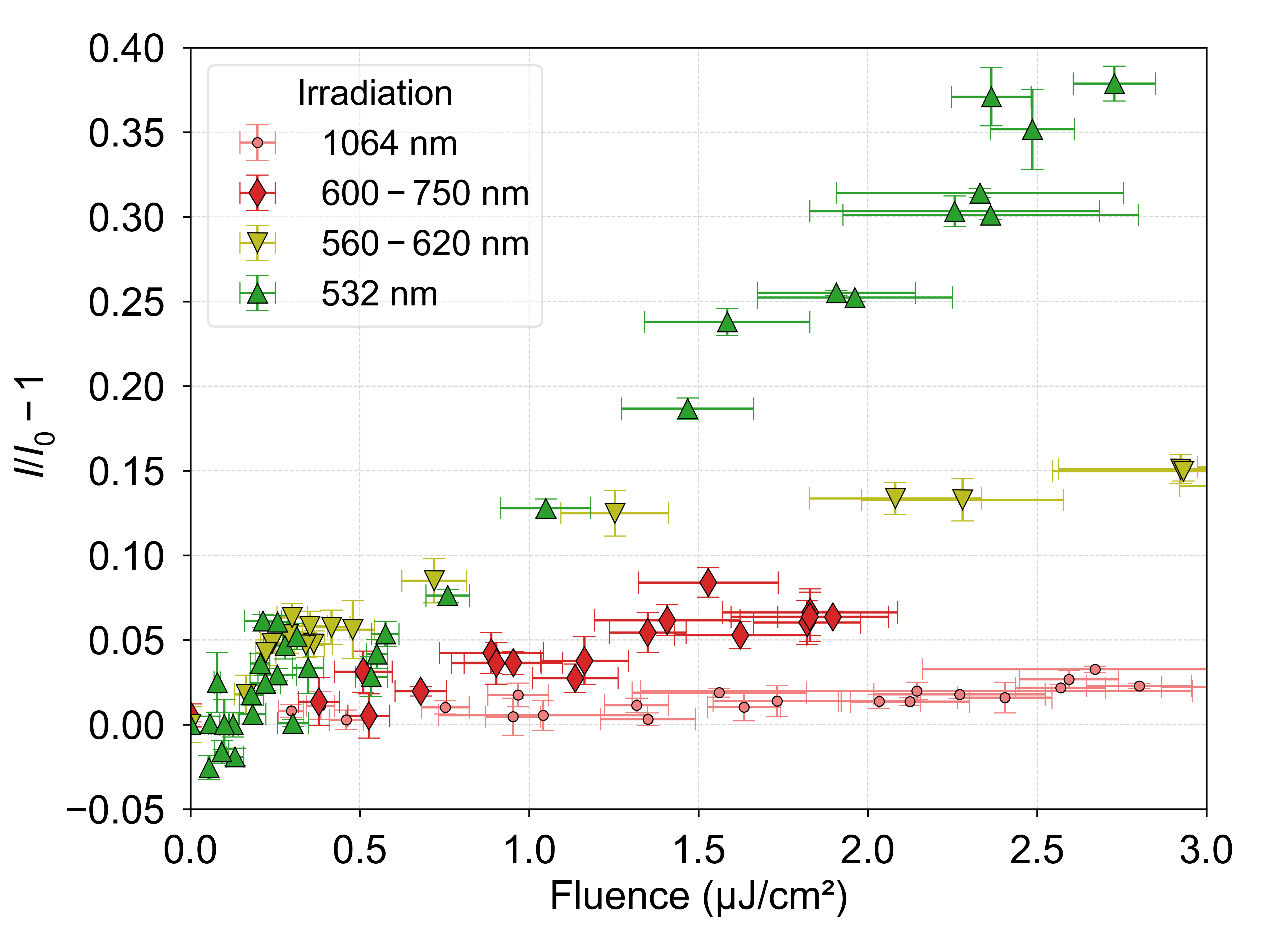}};
        
    \end{tikzpicture}

    \caption{\emph{Top:} Relative increase in plasma emission intensity induced by external illumination, as a function of illumination fluence per pulse. Individual frames were obtained with 400-ps exposure time and accumulated on-CCD 1000 times. Exposure began at $t_2$ = 7.3 ns. Errors bars represent the standard deviations over 2 to 10 frames per measurement, which were each repeated 2 to 4 times. \emph{Bottom:} Detailed view at low fluence of the highlighted region in the top panel.}
    
    \label{4emissionlarge}
\end{figure}

Above this threshold, the increase in plasma emission follows a log-linear behaviour as a function of illumination fluence :  

\begin{align}
\frac{I}{I_0} - 1 &= m \ln\!\left(F[\mathrm{J}/\mathrm{cm}^2]\right) + b \tag{1}
\label{eq:intensite}
\end{align}

where $m$ and $b$ are the slope and intercept of the relation, respectively. The slope of the increase is steeper for shorter wavelengths, indicating a stronger interaction. For instance, for 532 nm, $m = 0.197 \pm 0.007$, wheareas for 1064 nm, $m = 0.064 \pm 0.002$. The intercepts $b$ are $2.88 \pm 0.09$ and $0.83 \pm 0.02$ for 532 nm and 1064 nm, respectively.

While the plasma emission continues to increase according to Equation \ref{eq:intensite} up to the highest fluences applied here, above a certain fluence the enhancement extends beyond the irradiation area (Figure \ref{4laser} panel (d)). The enhancement "overflows" beyond the front of the ionization wave and into its core, at the thresholds of $F$ = 3.9 µJ/cm$^2$ for 532 nm and $F$ = 39.8 µJ/cm$^2$ for 1064 nm.

\subsection{Reduced electric field} \label{efield}

To study the effect of external irradiation on the reduced electric field, the irradiation timing and positioning were set as described in Section \ref{timeposirrad}. Figure \ref{3fente} illustrates the spatial overlap between the external illumination and the plasma front as viewed through the spectrometer entrance slit. The slit width was set to 20 µm in order to achieve sufficient spectral resolution for OES. However, this width was too narrow to permit a full view of the illuminated region. As a result, the measured FNS/SPS emission ratio corresponds only to a portion of the overlapping region.

\begin{figure}[!htbp]
\centering
\includegraphics[width=8cm]{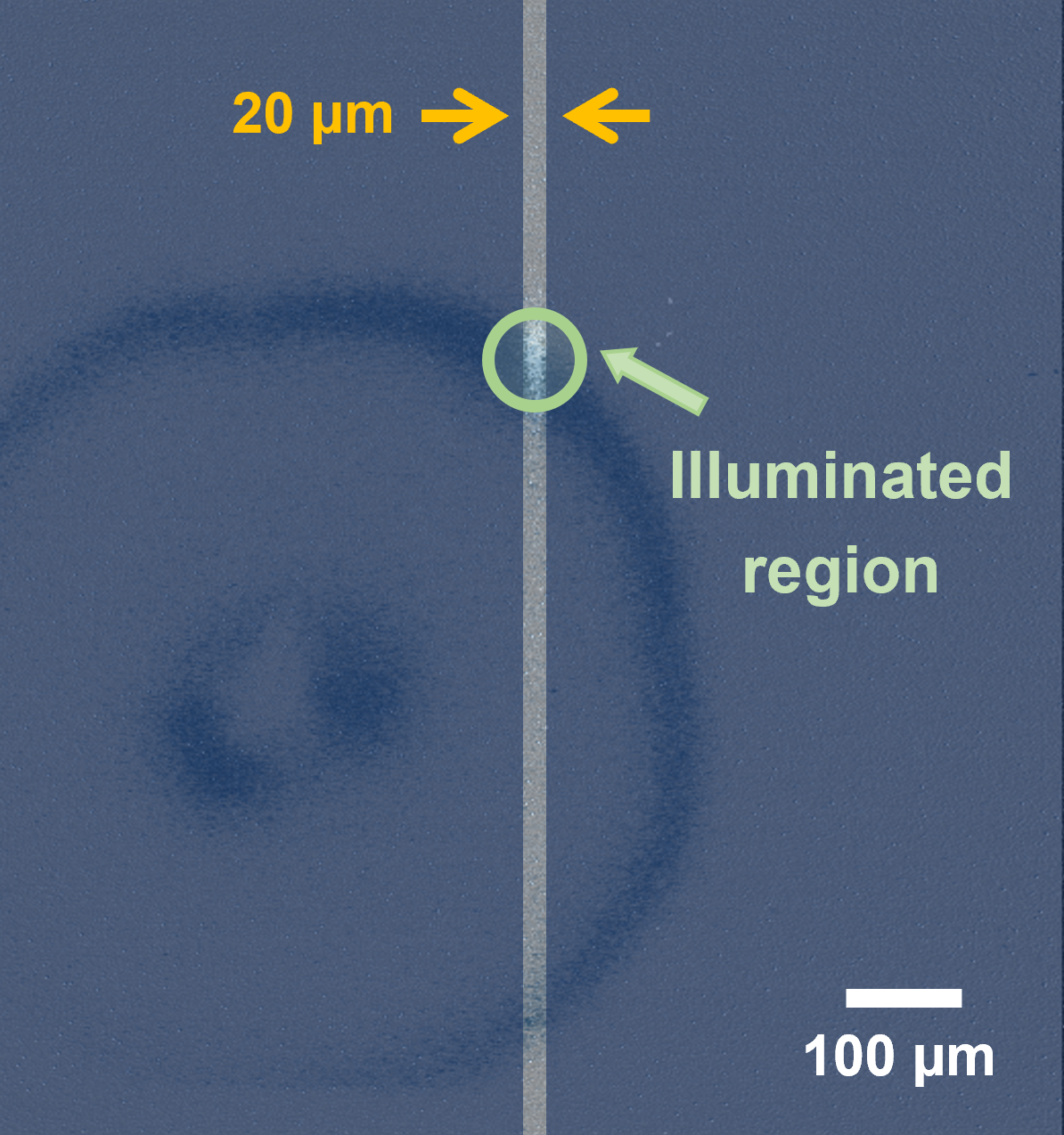} 

\caption{Negative of the entire discharge image at $t_2$, overlaid with the positive image of the 0.005 mm$^2$ light-exposed region viewed through the 20-µm entrance slit of the spectrometer.} 
\label{3fente}
\end{figure}

Figure \ref{4Eex} illustrates an example of $N_2^+~(B-X)~(0-0)$, $N_2^+~(B-X)~(2-1)$ and $N_2~(C-B)~(1-4)$ spectra with and without 532 nm irradiation at $F$ = 80 µJ/cm$^2$. Consistent with Figure \ref{4emissionlarge}, the overall intensity of the spectrum is higher than without irradiation. Moreover, the intensity of the N${_2^+}~(B-X)~(0,0)$ band increases more than that of N$_2~(C-B)~(1-4)$. 

\begin{figure}[!htbp]
\centering
\includegraphics[width=12cm]{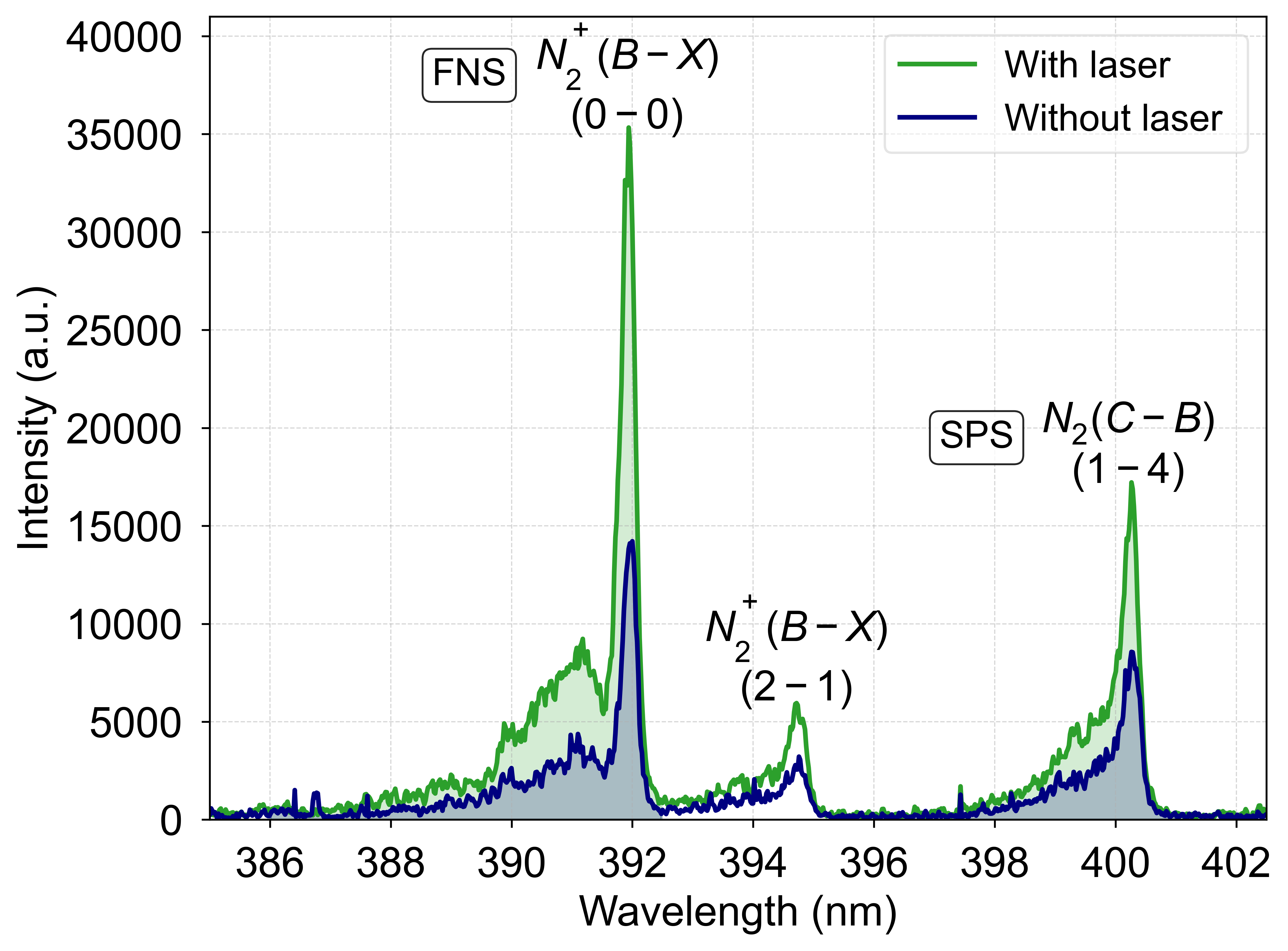} 

\caption{ Plasma emission spectra acquired at $t$ = 5.4 ns, without and with 532 nm laser irradiation at $F$ = 80 µJ/cm$^2$. The shaded regions are used to calculate the ratio $R$. These spectra were obtained using a 400 ps exposure time, accumulated over 2000 discharge events on CCD and averaged over 5 frames.} 
\label{4Eex}
\end{figure}

Figure \ref{4elecfield} illustrates the relative increase in the ratio of intensities of emission  $\frac{R - R_0}{R_0}$ from the $N_2^+~(B-X)~(0-0)$ versus $N_2~(C-B)~(1-4)$ bands as a function of the illumination fluence $F$ for 532 nm and 1064 nm. The intensity ratios $R$ and $R_0$ correspond to the cases with and without laser irradation, respectively. The reference value $R_0$ corresponds to the unperturbed plasma conditions shown in Figure \ref{3curve}.

\begin{figure}[!t]
\centering
\includegraphics[width=12cm]{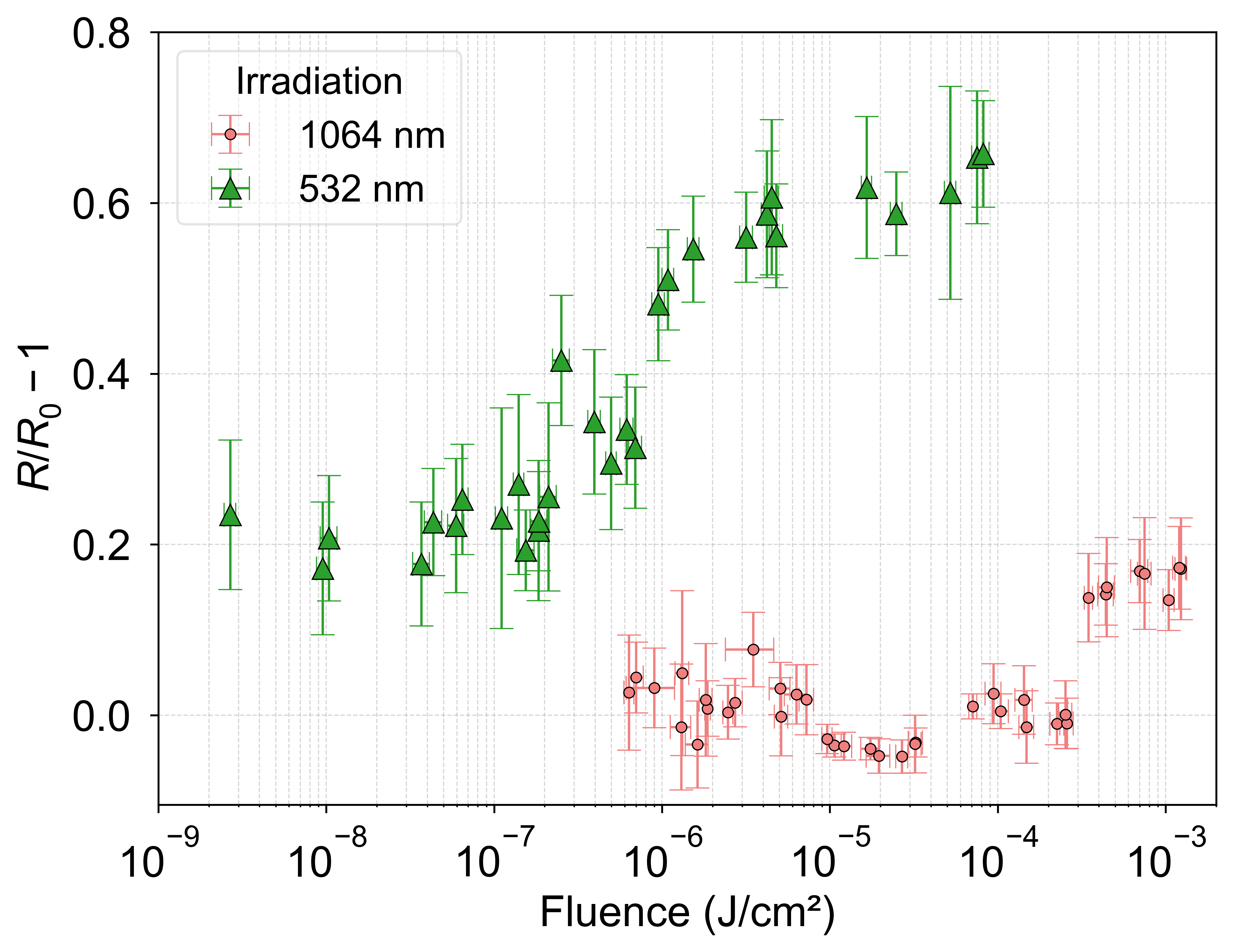} 

\caption{Relative increase of the intensity ratio of the $N_2^+~(B-X)~(0-0)$ to $N_2~(C-B)~(1-4)$ bands induced by external illumination, as a function of illumination fluence per pulse at two different wavelengths. Individual frames were obtained with 400-ps exposure time and accumulated on-CCD 2000 times.
Errors bars represent the standard deviations over 5 frames per measurement, which were each repeated 2 times.}
\label{4elecfield}
\end{figure}

For both laser wavelengths, the intensity ratio exhibits step-like increases. For 532 nm, at low fluences the intensity ratio is constant but increased by 20\% over $R_0$. Thus, even the lowest $F$ = 2.5 nJ/cm$^2$ results in a sizeable increase in $R$ and potentially the reduced electric field. The threshold fluence was below our detection limit but above zero. From $F$ = 0.1 to 4 µJ/cm$^2$, $\frac{R - R_0}{R_0}$ increases from 0.2 to 0.6 and remains at about this value at least up to $F$ = 90 µJ/cm$^2$. This transition reaches a midpoint values of 0.4 at about $F$ = 0.7 µJ/cm$^2$, which is also the threshold fluence for plasma emission enhancement shown in Figure \ref{4emissionlarge}. 
A plateau value is reached at 3 µJ/cm$^2$, coinciding with the onset of emission enhancement overflow discussed in the previous section (Figure \ref{4laser} (g)). In contrast, at 1064 nm, although plasma emission increases once $F$ exceeds 3 µJ/cm$^2$, $\frac{R - R_0}{R_0}$ remains unaffected by irradiation up to the threshold $F$ = 0.4 mJ/cm$^2$, above which it rises up to a constant value of 0.18 at least until $F$ = 1.3 mJ/cm$^2$ (Figure \ref{4elecfield}). Unlike for 532 nm irradiation, no transition to higher $\frac{R - R_0}{R_0}$ was observed.

\subsection{Discharge energy} \label{energy}

Figure \ref{4expuissance} presents the electric power curve as a function of time for $F$ = 1.3 mJ/cm$^2$ of irradiation at 1064 nm, showing a very low increase over the case without irradiation. The largest difference in power occurs at $t$ = 6 ns, which corresponds to the time during which the irradiation is active. At this moment, the peak power is 1.74 kW with irradiation compared to 1.62 kW without irradiation. This increase does not rise above uncertainty when averaged over 100 discharge events.

\begin{figure}[!htbp]
\centering
\includegraphics[width=12cm]{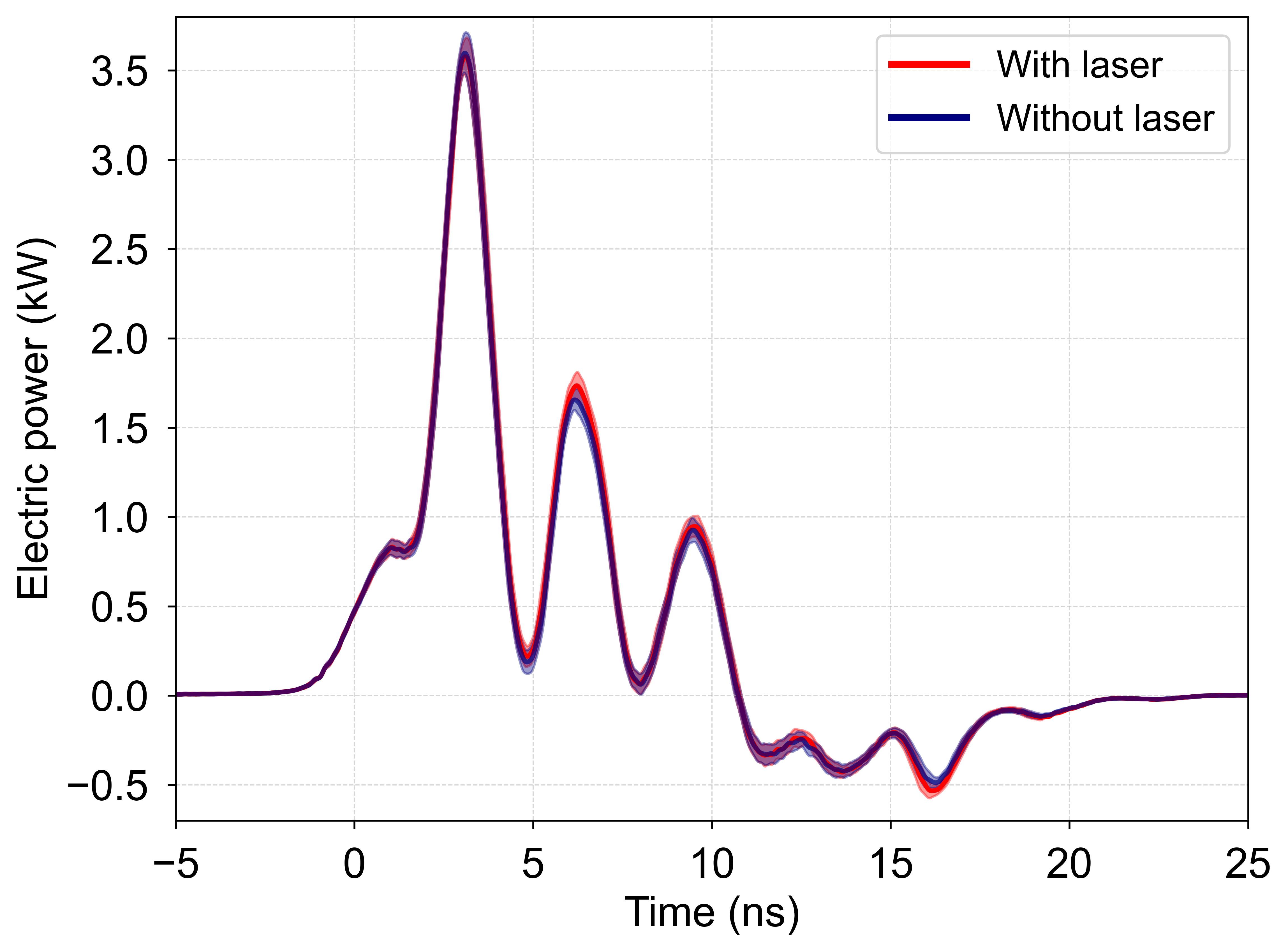} 

\caption{ Electric power without and with 1064 nm laser irradiation at $F$ = 1.3 mJ/cm$^2$. The waveforms were averaged over 100 discharge events.} 
\label{4expuissance}
\end{figure}

Figure \ref{4energylarge} illustrates the relative increase of discharge energy $\frac{U - U_{0}}{U_{0}}$ as a function of the illumination fluence $F$ for the same wavelengths shown in Figure \ref{4emissionlarge}. The energies $U$ and $U_{0}$ correspond to the cases with and without irradiation, respectively. The energy $U_{0}$ corresponds to the unperturbed plasma conditions shown in Figure \ref{3curve}.

\begin{figure}[!htbp]
\centering
\includegraphics[width=12cm]{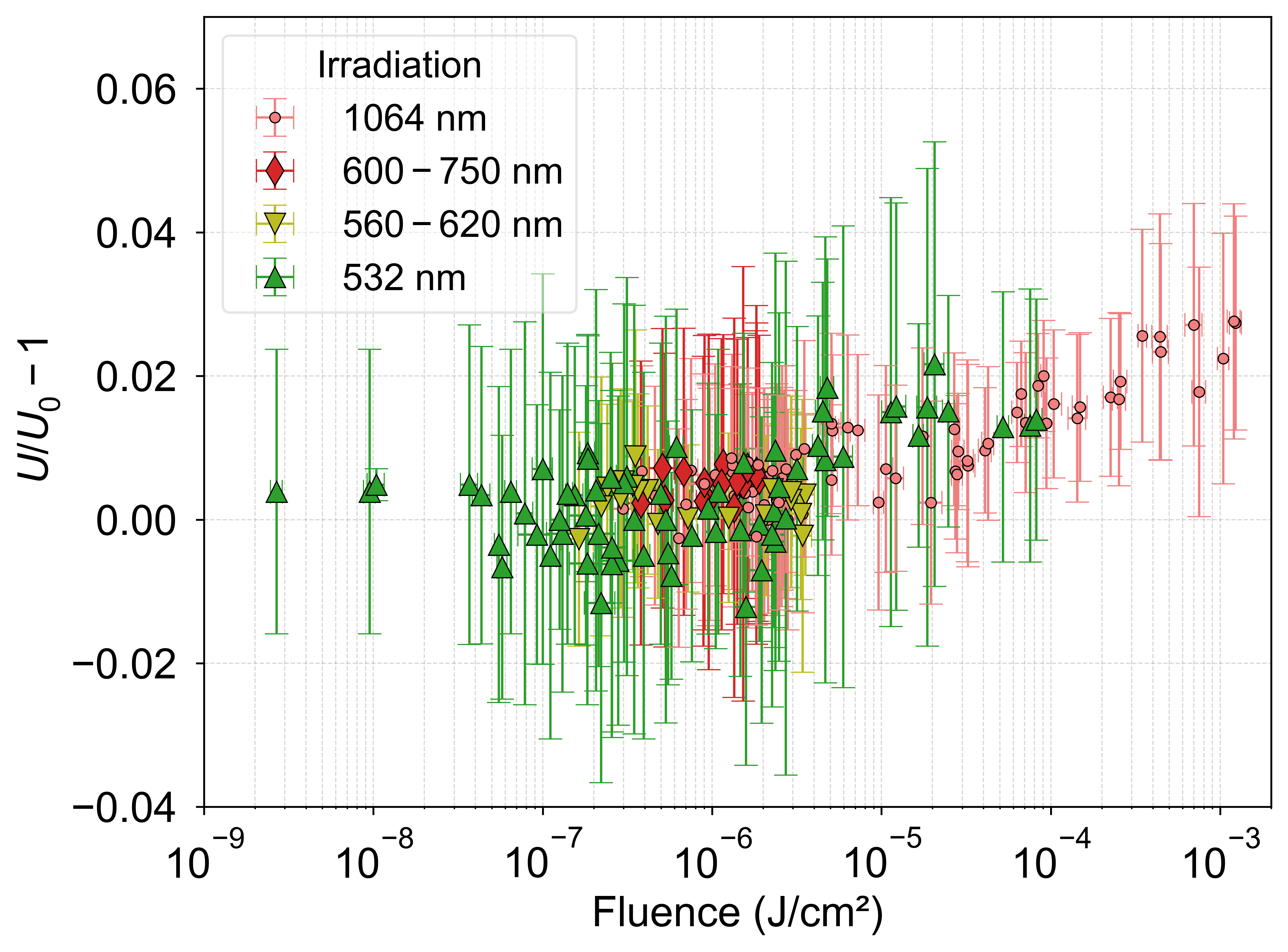}

\caption{Relative increase in discharge energy induced by external illumination, as a function of illumination fluence per pulse. Errors bars represent the standard deviation over 100 discharge events per measuremet, each of which was repeated 2 to 4 times.}
\label{4energylarge}
\end{figure}

For all wavelengths, the electrical energy does not exhibit any clear increase. It is possible that above a certain threshold of $F$ = 0.1 mJ/cm$^2$ for 1064 nm, the energy starts to rise. However, given the uncertainty, no definitive conclusion can be drawn up to $F$ = 1.3 mJ/cm$^2$ of irradiation. 
It is important to note that the laser energy is at most 6 nJ, while the SeBD electrical energy is on the order of 9 µJ, making the laser-induced energy perturbation negligible. Furthermore, the electrical energy measurement uncertainty (200 nJ) exceeds the laser energy by two orders of magnitude, rendering any direct photoconductive contribution undetectable.

\section{Discussion}

The study of plasma emission intensity and electric field shown in Figures \ref{4emissionlarge} and \ref{4elecfield} reveals that irradiation at shorter wavelengths produces a more pronounced effect on the plasma. To gain insight into this behaviour, we will discuss the interaction as it evolves in time. We will also propose mechanisms to explain the wavelength dependence and narrow down the range of relevant phenomena by process of elimination.

\subsection{Photons at the Air-SiO$_2$ interface} \label{airinterface}

We begin with the photon incidence at the air-SiO$_2$ interface. The Fresnel equations at two interfaces \cite{heavens1960optical} 
using the indices of refraction in \cite{palik1998handbook, malitson1965interspecimen}.
yield the reflection coefficients at normal incidence on the air-Si-SiO$_2$ interface of $\Gamma$ = 0.18 at 1064 nm (frequency $\nu_1$) and $\Gamma$ = 0.11 at 532 nm (frequency $\nu_2$), which are too close in value to explain the dependence on wavelength. Another effect to consider is the photodesorption of surface charge, which in principle reduces screening of the applied electric field and thereby increases plasma activity. Previous studies showed that photodesorption from dielectrics can be wavelength dependent \cite{ussenov2024laser, ziemba2025removal, guaitella2011role,tschiersch2017influence}. However, these previous studies \cite{ussenov2024laser, zhao2024repetitively} required fluences of around $F$ = 30 mJ/cm$^2$ at 532 nm to achieve significant desorption from surface DBDs. SeBDs produce much more current than surface DBDs yet extend over a smaller area. Therfore, we can expect that the surface charge density of SeBDs to be similar to that of surface DBDs, if not higher. It follows that the low fluences used in this study should desorb a negligible percentage of the surface charge. Furthermore, Orrière \textit{et al.} \cite{Orriere2026APL} demonstrated that changes in SeBD properties occur only when the delay of the 532 nm laser irradiation is less than 3 µs before plasma generation. The response of the plasma should not depend on the moment of photodesorption prior to breakdown, indicating that the laser energy is not being used to remove the space charge deposited by the plasma on the SiO$_2$. All these observations point to the wavelength dependence of the irradiation effects arising primarily from the silicon rather than the SiO$_2$ surface.

\subsection{Photons at the Si-SiO$_2$ interface and in the silicon bulk} \label{bulkpenetration}

An incident photon with energy $h\nu$, where $h = 6.63 \times 10^{-34}$ m$^2$.kg/s is Planck's constant and $\nu$ is the photon frequency, can initiate several processes in silicon. First, an electron in the valence band of silicon absorbs the photon if $h\nu > E_g = 1.12$ eV, up to a penetration depth $\delta$ governed by the Beer-Lambert law. 
The electron then transitions to the conduction band, leaving behind a hole. 
The excess energy $h\nu - E_g$ is dissipated as phonon excitation of the crystal lattice 
at the 100 ps timescale or less \cite{shah2013ultrafast} 
Phonon energy subsequently converts into thermal energy. In addition, the photogenerated free carriers can undergo recombination, diffusion, free carrier absorption (FCA) as well as drift and impact ionization in the presence of a high electric field. We will evaluate the contribution of each step to the observed results.

\subsubsection{Laser heating}

First, we consider laser heating of the silicon. In our study, the maximum possible excess energy that could be converted into thermal energy is supplied when using a wavelength of 532 nm at $F_{max} = 8.2 \times 10^{-5}$  J/cm$^2$ of irradiation per pulse. The excess energy available for absorption is $ Q_{abs} = (1-\Gamma) \times F_{max} \times S \times (1 - \frac{E_g}{h\nu_2}) = 1.9 \pm 0.5$ nJ, where $h\nu_2 = 2.33$ eV is the energy of the incident photon at 532 nm.
We calculate a maximum possible temperature increase of $ \Delta T = \frac{Q_{abs}}{VC_v} = 0.18 \pm 0.06$  K per pulse, where  $C_v = 1.66 \times 10^6 $ J/m$^3$ is the heat capacity of silicon at 300 K \cite{touloukian1971thermophysical} and $V = (6.39 \pm 1.60) \times 10^3$ µm$^3$ is the volume of laser absorption determined by a the laser spot area $S$ and $\delta$ = 1.27 µm, the absorption length in silicon at 532 nm \cite{green2008self}. This temperature rise is too small to affect any properties of silicon appreciably, so laser heating cannot account for the differences observed in our results.

\subsubsection{Comparison with a MOS photodetector} \label{sec:mos}

As a first approximation, the irradiation of the SeBD may function similarly to a Metal-Oxide Semiconductor (MOS) photodetector. We will consider the positive SeBD as a positively biased metal electrode. In the strong inversion regime for these devices, a few volts are applied over an oxide layer of a few nanometers \cite{hu2010modern}.
In the case of the SeBD, typically a few kilovolts are applied over 1-µm of oxide, and by proportionality we can therefore expect the formation of both an inversion region of $1-3$ nm thickness \cite{hu2010modern} with a high density of free electrons and a depletion region of $0.3-0.9$ µm for the doping level under consideration \cite{sze2021physics} in which holes are absent, leaving only the fixed negatively charged dopant atoms ($B^-$). Such a charge structure is illustrated by Figure \ref{5impact532}. This creates a strong electric field within the depletion region, leading to the efficient separation of photoexcited electron-hole pairs. The electric field is low in the rest of the bulk silicon due to screening by the inversion region and depletion layers. Similar structures are employed in MOS photodetectors, where devices are designed such that absorption occurs in the depletion region. 
Similarly for the SeBD, the effect of photoexcited carriers can be amplified depending on the penetration depth in relation to the likely spatial structure of charge in the silicon in relation to the penetration depth.

The penetration depth $\delta$ for intrinsic silicon is 1.27 µm for 532 nm irradiation, $1.73 - 2.84$ µm for $560 - 620$ nm irradiation, $2.40 - 7.36$ µm for $600 - 750$ nm irradiation, and 0.9 mm for 1064 nm irradiation \cite{green2008self}. For moderately doped silicon, the energy band structure remains largely unchanged \cite{sze2021physics}, so it is reasonable to assume absorption coefficients similar to intrinsic silicon. 
Therefore, we can expect that with 532 nm irradiation, the penetration depth falls within the depletion region (Figure \ref{5impact532}), whereas at 1064 nm irradiation most of the photoexcitation occurs beyond this zone (Figure \ref{5impact1064}), making electron-hole pair separation less efficient.
As a consequence, shorter-wavelength excitation leads to the generation of a higher density of electron-hole pairs closer to the interface. Such a high-density carrier region close to the surface can enhance electric field interaction with the air plasma and lower the fluence threshold for ionization wave perturbation at 532 nm compared to 1064 nm. This provides a primary factor explaining how irradiation wavelength influences the fluence threshold shown in Figure \ref{4emissionlarge} or equivalently the intercept $b$ from Equation \ref{eq:intensite}.

Furthermore, the observed irradiation fluences for the thresholds for enhancing plasma emission intensity and for the transition of electric field enhancement suggest that a critical density of electron-hole pairs is necessary to create these effects. At 532 nm, both effects share a common threshold/transition $F$ = 0.7 µJ/cm$^2$, which corresponds to $9.4 \times 10^7$ photons absorbed in $V$ after accounting for 89\% transmission through the SiO$_2$-Si interface. Assuming a photogeneration probability of 100\% leads to a carrier density of $p = 8.3 \times 10^{15}$ cm$^{-3}$. This value is of the same order of magnitude as the equilibirum hole density of  $p_0 = (10^{15} - 2 \times 10^{16})$ cm$^{-3}$ for $1 - 10~\Omega\cdot$ cm p-doped silicon at 300 K, suggesting that the strongest perturbation of the SeBD requires the injection of $\approx p_0$ free carriers, which is the same density as $B^-$ atoms in the depletion region.
In contrast, at 1064 nm, the threshold for emission enhancement is $F$ = 3 µJ/cm$^2$ corresponding to $8 \times 10^8$ photons. Given 82\% transmission through the SiO$_2$-Si interface and penetration through the entire silicon volume $V$ = 2.64 $\times$ 10 $^{-3}$ mm$^3$, the resulting carrier density is $p = 2.5 \times 10^{14}$ cm$^{-3}$, which is only $1.25-12.5$ \% of $p_0$. Thus, the ionization wave responds to a small increase in charge carrier density throughout the silicon, likely through a bulk interaction mechanism that differs from the interfacial interaction induced by 532 nm irradiation.
It follows that the mechanisms governing SeBD perturbation require the injected free carriers to at least approach the density $p_0$ and also depend on the proximity to the SiO$_2$-Si interface.

\begin{figure}[!htbp]
\centering
\includegraphics[width=10cm]{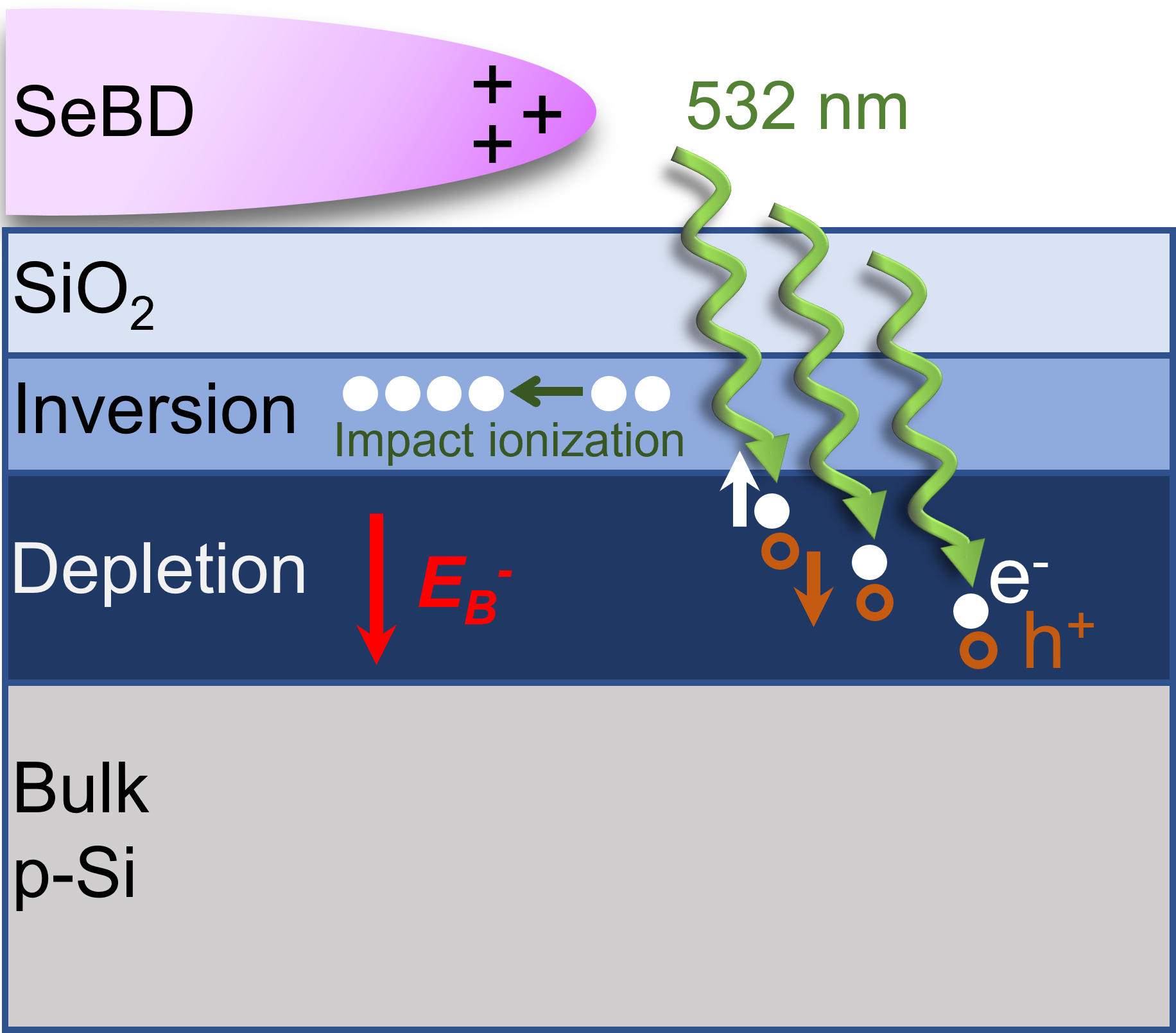} 

\caption{Schematic illustration of photoexcited carrier generation and separation followed by impact ionization, near the SeBD-SiO$_2$-Si interface under 532 nm illumination. Note that depending on the stregth of the total electric field (generated by the gas-phase plasma in the silicon and $\boldsymbol{E_{B^-}}$ generated by the $B^-$ atoms), impact ionization can also take place in the depletion region.} 
\label{5impact532}
\end{figure}

\begin{figure}[!htbp]
\centering
\includegraphics[width=10cm]{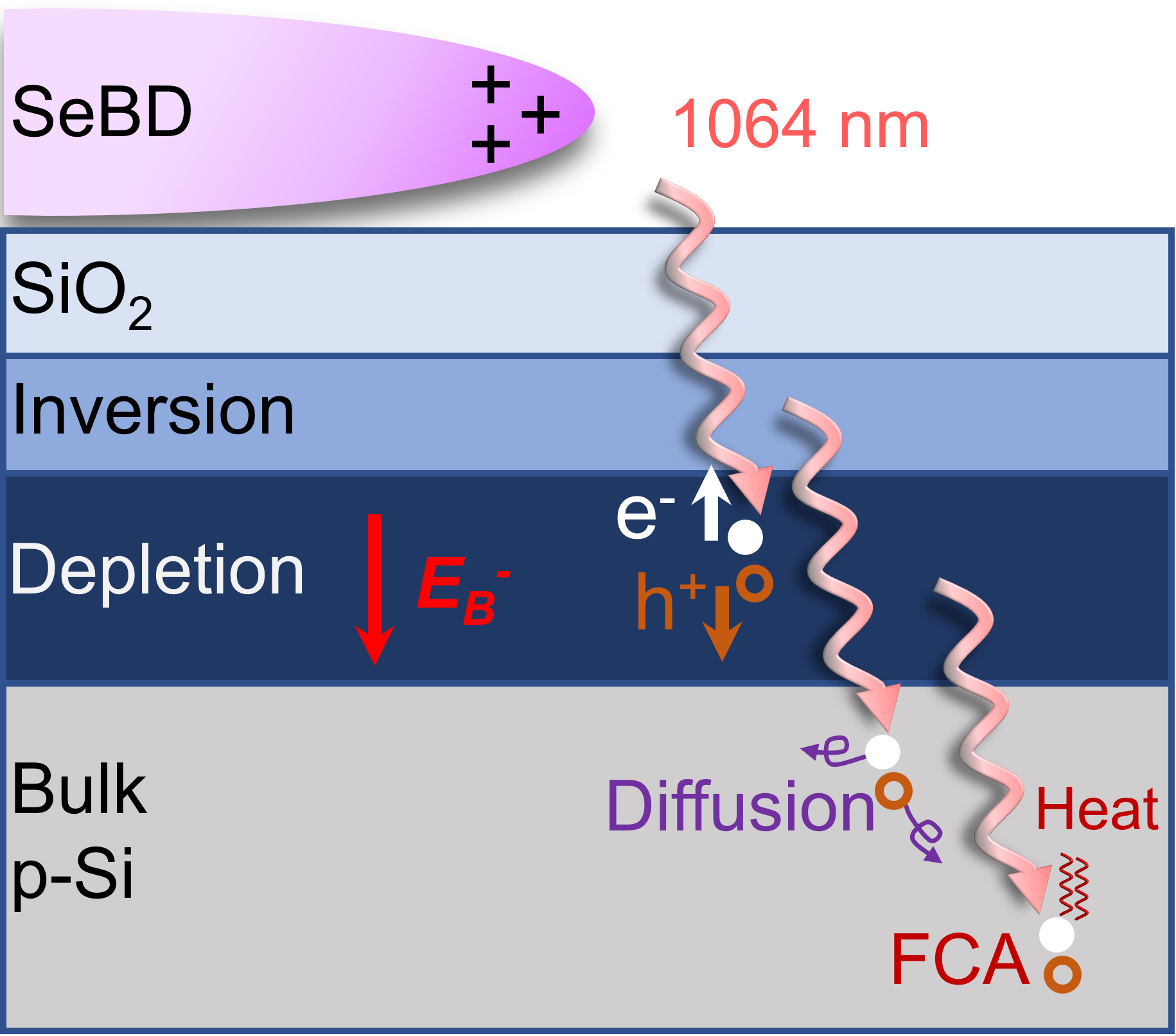} 

\caption{Schematic illustration of photoexcited carrier generation and separation, followed by FCA and diffusion throughout the bulk silicon under 1064 nm illumination. Note that FCA can also occur in the inversion region.}

\label{5impact1064}
\end{figure}

In addition, the wavelength dependence of the slope $m$ from Equation \ref{eq:intensite} points to yet other specific processes occurring within the silicon. 
We suppose the amount of enhancement of plasma emission intensity to be linearly proportional to the yield of some process in the silicon induced by irradiation. For example, Figure \ref{4emissionlarge} shows that an enhancement of 0.2 requires  $F$ = 100 µJ/cm$^2$ at 1064 nm compared to less than $F$ = 2 µJ/cm$^2$ pJ at 532 nm. Supposing that the enhancement is directly proportional to the number of photoexcited free carriers, we can equate the log-linear fits to the data from Figure \ref{4emissionlarge} to obtain the relation $\epsilon_{2}^{3} \propto \epsilon_1$, where $\epsilon_1$ and $\epsilon_2$ are the irradiation energies at 1064 and 532 nm, respectively. This implies that producing the same number of free carriers requires the equivalent of three photons at 1064 nm for every single photon at 532 nm : 

\begin{align}
h\nu_{2} &\xrightarrow{\text{Si}} e^- + h^+
\qquad\Leftrightarrow\qquad
3\,h\nu_{1} \xrightarrow{\text{Si}} e^- + h^+  \tag{2}
\label{eq:correspondance}
\end{align}

Multiple processes may contribute to the overall balance of photons and charge carriers represented by process \ref{eq:correspondance}. To understand how such a situation could arise, we consider the quantum efficiency (QE) of photoexcitation at these two wavelengths. First, the QE can decrease because of FCA, where free electrons/holes already in the conduction/valence band absorb photons. FCA thus diverts irradiation energy toward heating existing free carriers instead of generating new carriers. FCA features prominently at 1064 nm but is negligible at 532 nm \cite{lietoila1982computer}. Second, the QE can increase due to impact ionization, whereby a photoexcited free carrier acquires sufficient energy via the electric field to create an electron-hole pair, in a manner analogous to gas-phase electron-impact ionization \cite{sze2021physics}.
This effect strengthens when photoexcitation results in "hot" carriers with excess energy $h\nu-E_g$ that can boost impact ionization under circumstances when this energy loss channel becomes more probable than phonon excitation. UV irradiation pushes QE above 100\% \cite{kolodinski1993quantum}, which means that irradiation by the plasma emitting in the UV via the FNS and SNS transitions is capable of effectively producing more than one electron-hole pair per photon. The external irradiation used here is not energetic enough to produce QE greater than 100\% on its own, but hot carriers can acquire additional energy through the electric field \cite{tam1984hot} 
of the SeBD. Considering that $E_g$ = 1.12 eV, the excess energy for irradiation at 532 nm is $h\nu_2-E_g = 1.21$ eV compared to $h\nu_1-E_g = 0.05$ eV at 1064 nm. This means that electric field-assisted impact ionization of hot carriers is more likely at 532 nm than 1064 nm, especially considering that the threshold energy for ionization is 1.12 eV \cite{sano1992impact}. Thus, FCA can diminish QE at 1064 nm while hot-carrier impact ionization can increase QE at 532 nm, which could contribute to the observed 3:1 photon consumption ratio expressed by Equation \ref{eq:correspondance}. Furthermore, FCA should occur mostly in the Si bulk in the case of irradiation at 1064 nm (Figure \ref{5impact1064}), whereas impact ionization should be localized in the depletion region for irradiation at 532 nm (Figure \ref{5impact532}).

\subsubsection{Charge transport in silicon}

The strong interfacial electric field in the silicon enables electrons to follow the ionization wave. The typical electric field in silicon close to the saturation regime is approximately $E \approx 10^4$ V/cm \cite{sze2021physics}. In the saturation regime of transport, the drift velocity of charge carriers reaches an upper limit of about $10^5$ m/s due to increased scattering mechanisms, mainly with optical phonons. This velocity saturation occurs just before the onset of avalanche breakdown. During a 2-ns long laser pulse, 
charge can drift about $L \approx$ 200 µm at the saturation velocity, a significant fraction of the discharge extent. This implies that when external illumination generates electron-hole pairs, these charge carriers drift at a similar timescale as ionization wave propagation (Figure \ref{5impact532}). 
For the same characteristic length $L$ = 200 µm, the diffusion time is t$_{diffusion}$ =  $\frac{L^2}{2D} \approx$ 5.3 µs for electrons and 15.3 µs for holes, where the diffusivity in silicon at 300 K is $D = 3.75 \times 10^{-3}$ m$^2$/s for electrons and $1.31 \times 10^{-3}$ m$^2$/s for holes. This is much larger than the irradiation and plasma durations. 
Thus, drift dominates over diffusion in the interfacial high electric field region. However, in the quasi-neutral bulk silicon region where the electric field is weak, diffusion becomes more dominant than drift (Figure \ref{5impact1064}). 

The charge transport mechanism determines whether photogenerated charge carriers can separate efficiently. In MOS photodetectors, efficient charge separation enhances carrier collection and, consequently, the photocurrent \cite{holst2007cmos}. Drift is more effective than diffusion at separating charge because carriers are driven by the built-in electric field in the depletion region, which rapidly sweeps electrons and holes in opposite directions. Therefore, irradiation at 532 nm, which is strongly absorbed near the surface and within the depletion region, results in efficient charge separation and collection, leading to a high photocurrent response. On the other hand, irradiation at 1064 nm, which penetrates deeply into bulk silicon where the electric field is weak or absent, results in poor charge separation dominated by diffusion and consequently a reduced photocurrent.

\subsubsection{Recombination and memory effect}

During the post-discharge phase, four types of recombination \cite{plakhotnyuk2018nanostructured} are important to take into account in silicon : radiative and Auger band-to-band recombination (which are intrinsic), but also Shockley-Read-Hall (SRH) \cite{shockley1952statistics} bulk and surface recombination which are trap-assisted (extrinsic) processes. In radiative recombination, an electron from the conduction band recombines with a hole in the valence band, emitting a photon \cite{sze2021physics}. This process is fast in direct bandgap semiconductors. In contrast, silicon has an indirect bandgap, meaning that electron-hole recombination requires the participation of a phonon to conserve momentum, slowing the process considerably. For the excess charge carrier densities induced by external irradiation in this work ($p = 10^{14} - 10^{17}$ cm$^{-3}$), the radiative recombination lifetime in silicon is estimated to be $1-6$ ms \cite{macdonald2001recombination, schroder2015semiconductor}. Auger recombination \cite{sze2021physics} 
involves three carriers: an electron and a hole recombine transferring energy to a third carrier, typically an electron in the conduction band. This electron then thermalizes back to the conduction band edge. Auger recombination becomes significant in heavily doped semiconductors. For the doping levels and excess carrier densities under consideration, its characteristic time is in the range of 50 µs $-$ 5 ms \cite{macdonald2001recombination, schroder2015semiconductor, rein2005lifetime}. Third, in SRH recombination, a carrier is captured by a trap state with an energy level within the band gap originating from crystal defects or impurities. Trap times decrease with increasing electric field \cite{kramberger2002effective} 
and strongly depend on the type and concentration of defects \cite{kramberger2002determination}. Bulk SRH recombination \cite{macdonald2001recombination, schroder2015semiconductor} occurs at the 0.2 $-$ 2 ms time scale. Interface traps in silicon are not characterized by a single relaxation time but instead exhibit a broad spectrum of capture and emission dynamics on timescales ranging from microseconds \cite{tan2020facet} 
to minutes \cite{grasser2010time, kumar2024exploring}. 
All the aforementioned time scales are significantly longer than the pulse duration but remain shorter than the interpulse period of 10 ms, except for the interface trap lifetime.

On this basis, although difficult to quantify in our study, SRH recombination at the SiO$_2$-Si interface could nonetheless contribute to the memory effect shown in panel (h) (Figure \ref{4laser}) because its trap lifetime is the only potentially long enough to influence subsequent discharges.
The memory effect is unlikely to result from permanent surface modification, since the plasma extension gradually returns to its initial state after a few seconds or minutes. The reversibility of this process instead points toward transient mechanisms, such as SRH surface recombination. Thermal effects are negligible, as previously discussed. 

The memory effect of DBDs is often attributed to residual surface charge, which can have long lifetimes \cite{opaits2008surface}. In our study, pulsed irradiation may locally increase ionization by the air plasma through the coupling mechanism proposed in Section \ref{sec:mos}, which could remain undetected for the reasons to be presented in Section \ref{sec:energydisc} explaining why the energy remains unchanged with irradiation intensity (Figure \ref{4energylarge}). In turn, this could result in increased local deposition of surface charge and therefore increased screening of the applied electric field. At the highest fluences, the plasma alone may not be able to overcome the increased screening, thus inducing the memory effect of the observed plasma "void" after stopping irradiation. Moreover, there was no additional irradiation between discharges to perform desorption of the surface charge and alleviate screening. According to this reasoning, additional charge generation would have also occurred in the irradiation experiments of Darny \textit{et al.} \cite{darny2020uniform}. However, because the laser was continuous in that case, there was a means to desorb surface charge between pulsed discharges. Yet, Darny \textit{et al.} \cite{darny2020uniform} still observed the memory effect in the same manner as this study. 

Furthermore, comparing the irradiances between Darny \textit{et al.} \cite{darny2020uniform} and our study shows that these observations are consistent with a memory effect associated with the trapping of charge but not with increased ionization by the air plasma. Darny \textit{et al.} \cite{darny2020uniform} used continuous wave laser irradiation at an irradiance $I$ = 60 W/cm$^2$, which was sufficient to produce the memory effect. In our study with pulsed irradiation, for the fluence threshold of $F$ = 25 µJ/cm$^2$ for the memory effect shown in panel (g) of Figure \ref{4laser}, the instantaneous irradiance during the laser pulse is $I$ = 10 kW/cm$^2$, and the average irradiance is $I$ = 2 mW/cm$^2$. Increased ionization by the air plasma upon irradiation should depend on the instantaneous irradiance because both the plasma and the laser must be present to create this effect. However, the instantaneous irradiance used by Darny \textit{et al.} \cite{darny2020uniform} is much lower than the threshold found here, which would indicate that not enough additional charge was generated during the plasma to produce the observed memory effect. On the other hand, the impact of trapped charge should depend on the average irradiance because the long lifetime of traps, presumably determined by SRH surface recombination, results in a cumulative effect. The average irradiance used by Darny \textit{et al.} \cite{darny2020uniform} is higher than the threshold determined in our study, and therefore a memory effect induced by trapped charge is consistent with all reported irradiation experiments.

\subsection{Discharge electrical energy} \label{sec:energydisc}

Finally, several mechanisms may account for the observed local enhancement of plasma emission and electric field (Figures \ref{4emissionlarge} and \ref{4elecfield}) without any change in electrical energy (Figure \ref{4energylarge}). First, the measured energy may not be dissipated only in the gas-phase plasma but also within the silicon. Some decoupling may be expected between the total energy and the properties of the air plasma. Second, the region of enhanced plasma emission and electric field is small compared to the total size of the discharge. A local increase in discharge energy may be too small in proportion to rise above the measurement uncertainty. Another possibility is that the energy does not need to vary because the increased plasma emission at the location of irradiation is compensated by decreased plasma emission elsewhere along the ionization wave front, which can be observed by comparing panels (g) and (f) to (e) of Figure \ref{4laser}. At laser fluences higher than used in this study, the energy does increase upon irradiation in Orrière \textit{et al.} \cite{Orriere2026APL}.

\vspace{-0.1cm}

\section{Conclusion}

This work has investigated the photonic aspect of how a surface ionization wave couples to a semiconductor. External illumination, ranging from 532 nm to 1064 nm and delivering up to several nanojoules per light pulse to the semiconductor barrier discharge (SeBD), was found to enhance both plasma emission and the reduced electric field, while leaving the discharge energy unaffected. These behaviours are strongly dependent on the irradiation fluence and wavelength, with shorter wavelengths producing a more pronounced effect. While the relative plasma emission rises steadily in a log-linear manner with increasing fluence, the electric field increases in steps. The overall trends remain similar from 532 nm to 1064 nm, but the thresholds for both plasma emission and electric field enhancement change significantly with wavelength, as does the rate of increase in plasma emission with fluence. 

These differences primarily originate from the wavelength-dependent absorption length and charge carrier processes in silicon. Shorter penetration depths at shorter wavelengths localize photoexcitation of electron-hole pairs closer to the SiO$_2$-Si interface. By comparing the SeBD to MOS photodetectors, two perturbation mechanisms can be distinguished. The first mechanism is interfacial and occurs within the depletion region under 532 nm irradiation. At this wavelength, absorption takes place inside an "amplification zone", where hot carriers and the strong electric field combine to reinforce impact ionization. This amplification is sufficient to induce gas-phase electric-field enhancement, detectable even at the lowest applied fluences. Upon increasing the fluence, a threshold is reached when the photoexcited carrier density equals the equilibrium charge carrier density $p_0$, triggering both an increase in plasma emission intensity and further electric-field enhancement. Above a fluence $F$ = 3.9 µJ/cm$^2$, the electric-field enhancement saturates, and the spatial extent of the enhanced plasma emission expands beyond the irradiated spot. The second mechanism is a volume effect occurring throughout the quasi-neutral silicon bulk under 1064 nm irradiation. At this wavelength, photoexcitation competes with free carrier absorbtion (FCA), while impact ionization remains negligible because the electric field in silicon is low beyond the depletion region. A first threshold of $F$ = 3 µJ/cm$^2$ corresponds to a rise in plasma emission intensity alone, which extends beyond the irradiated area starting at $F$ = 39.8 µJ/cm$^2$.

Several key questions remain unresolved. One key unknown property is the spatio-temporal distribution of the electric field inside silicon during the discharge. Another challenge lies in understanding the absence of a direct correlation between plasma emission and its electric field at 1064 nm, in contrast with the clearer relationship established at 532 nm. Additionally, the role of interface traps in the long-timescale memory effect observed at high fluence was difficult to fully establish.

To better understand these phenomena, additional diagnostics could be employed in future studies, such as in-situ Raman spectroscopy to characterize the SiO$_2$-Si interface \cite{pai2019situ} and shorter-duration laser pulses at wavelengths below 532 nm to more accurately mimic the emission spectrum of the air plasma. Beyond the semiconductor processes and device physics considered in this work, numerical modelling that accounts for the dynamics in the gas and solid phases is currently underway.

\vspace{0.3cm}

\subsection*{Acknowledgments}

We acknowledge the valuable support from MESR, EUR PLASMAScience, PEPS INSIS PlasmaTwins, Fédération Plas@Par, and ANR JCJC PLASMAFACE (ANR-15-CE06-0007-01).

\clearpage

\bibliographystyle{unsrt}
\bibliography{Biblio.bib}

\end{document}